\documentclass[journal,romanappendices]{IEEEtran}
\usepackage{algorithm,algorithmic,amsmath,amssymb,amsthm,bbm,cite,color,comment,enumitem,graphicx,microtype,subcaption,setspace,url}
\usepackage{mathtools,xparse}
\usepackage{diagbox}
\usepackage[USenglish]{babel}
\usepackage{hyperref}
\usepackage[utf8]{inputenc}
\usepackage[T1]{fontenc}
\usepackage{float}
\usepackage[autostyle]{csquotes}
\usepackage[labelsep=period,font=small]{caption}
\usepackage[nolist]{acronym}
\usepackage[normalem]{ulem}

\usepackage{tikz,pgfplots}
\usetikzlibrary{arrows, arrows.meta, backgrounds, calc, intersections, decorations.markings, decorations.pathreplacing, positioning, shapes}
\pgfplotsset{compat=1.17}
\newcommand{\Cross}{\mathbin{\tikz [x=1.4ex,y=1.4ex,line width=.2ex] \draw (0,0) -- (1,1) (0,1) -- (1,0);}}%
\usepackage{caption}
\usepackage{acronym}

\renewcommand{\a}{\mathbf{a}}

\newcommand{\e}{\mathbf{e}}

\renewcommand{\j}{\mathbf{j}}
\renewcommand{\k}{\mathbf{k}}

\renewcommand{\r}{\mathbf{r}}
\newcommand{\s}{\mathbf{s}}

\newcommand{\x}{\mathbf{x}}
\newcommand{\y}{\mathbf{y}}
\newcommand{\z}{\mathbf{z}}

\newcommand{\0}{\mathbf{0}}

\newcommand{\A}{\mathbf{A}}

\renewcommand{\H}{\mathbf{H}}
\newcommand{\I}{\mathbf{I}}

\newcommand{\R}{\mathbf{R}}

\newcommand{\W}{\mathbf{W}}

\newcommand{\gammab}{\boldsymbol{\gamma}}

\newcommand{\kappab}{\boldsymbol{\kappa}}

\newcommand{\sigmab}{\boldsymbol{\sigma}}

\newcommand{\psib}{\boldsymbol{\psi}}

\newcommand{\Sigmab}{\mathbf{\Sigma}}

\newcommand{\setD}{\mathcal{D}}
\newcommand{\setE}{\mathcal{E}}

\newcommand{\setL}{\mathcal{L}}

\newcommand{\setN}{\mathcal{N}}

\newcommand{\setS}{\mathcal{S}}
\newcommand{\setT}{\mathcal{T}}

\newcommand{\setW}{\mathcal{W}}

\newcommand{\Compl}{\mbox{$\mathbb{C}$}}
\newcommand{\Real}{\mbox{$\mathbb{R}$}}

\newcommand{\diag}{\mathrm{diag}}

\newcommand{\diff}{\mathrm{d}}
\newcommand{\Exp}{\mathbb{E}}
\newcommand{\herm}{\mathrm{H}}

\newcommand{\tr}{\mathrm{tr}}
\newcommand{\tran}{\mathrm{T}}

\renewcommand{\vec}{\mathrm{vec}}
\newcommand{\card}{\mathrm{card}}

\newtheorem{remark}{Remark}

\definecolor{oulu_blue}{HTML}{23408F}
\definecolor{oulu_green}{HTML}{39B54A}

\definecolor{red}{rgb}{1,0,0}
\definecolor{red_magenta}{rgb}{1,0,0.5}
\definecolor{magenta}{rgb}{1,0,1}
\definecolor{blue_magenta}{rgb}{0.5,0,1}
\definecolor{blue}{rgb}{0,0,1}
\definecolor{blue_cyan}{rgb}{0,0.5,1}
\definecolor{cyan}{rgb}{0,1,1}
\definecolor{green_cyan}{rgb}{0,1,0.5}
\definecolor{green}{rgb}{0,1,0}
\definecolor{green_yellow}{rgb}{0.5,1,0}
\definecolor{yellow}{rgb}{1,1,0}
\definecolor{red_yellow}{rgb}{1,0.5,0}
\begin{acronym}
\acro{1D}{one-dimensional}
\acro{2D}{two-dimensional}
\acro{3D}{three-dimensional}
\acro{ACF}{autocorrelation function}
\acro{AWGN}{additive white Gaussian noise}
\acro{DoF}{degrees of freedom}
\acro{EM}{electromagnetic}
\acro{i.i.d.}{independent and identically distributed}
\acro{ITU}{International Telecommunication Union}
\acro{LoS}{line-of-sight}
\acro{MIMO}{multiple-input multiple-output}
\acro{NLoS}{non-line-of-sight}
\acro{OFDM}{orthogonal frequency-division multiplexing}
\acro{PSD}{power spectral density }
\acro{PSF}{power spectral factor}
\acro{SNR}{signal-to-noise ratio}
\acro{vMF}{von Mises-Fisher}
\acro{WDM}{wavenumber-division multiplexing}
\acro{CSI}{channel state information}
\acro{TCA}{tightly coupled array}
\acro{LWA}{leaky-wave antennas}
\end{acronym}

\newcommand{\los}{\textnormal{LoS}}
\newcommand{\nlos}{\textnormal{NLoS}}

\title{LoS+NLoS Holographic MIMO: Analysis and Application of Wavenumber-Division Multiplexing}
\author{Ashutosh Prajapati, Prathapasinghe Dharmawansa, Marco Di Renzo,~\IEEEmembership{Fellow, IEEE}, \\ and Italo Atzeni,~\IEEEmembership{Senior Member, IEEE}
\thanks{A.~Prajapati, P.~Dharmawansa, and I.~Atzeni are with the Centre for Wireless Communications, University of Oulu, Finland (e-mail: \{ashutosh.prajapati, prathapasinghe.kaluwadevage, italo.atzeni\}@oulu.fi).}
\thanks{M. Di Renzo is with CNRS and CentraleSup\'elec - Univ. Paris-Saclay, Institute of Electronics and Digital Technologies (IETR), Avenue de la Boulaie, 35576 Cesson-S\'evign\'e, France (marco.direnzo@centralesupelec.fr), and with King's College London, Department of Engineering-Centre for Telecommunications Research, WC2R 2LS London, United Kingdom (marco.di\_renzo@kcl.ac.uk).}
\thanks{This work was presented in part at ASILOMAR 2025 \cite{Praj25}.}
\thanks{The work of A.~Prajapati, P.~Dharmawansa, and I.~Atzeni was supported by the Research Council of Finland (336449 Profi6, 348396 HIGH-6G, and 369116 6G~Flagship). The work of M. Di Renzo was supported in part by the Nokia Foundation, the French Institute of Finland, the French Embassy in Finland under the France-Nokia Chair of Excellence in ICT, and the European Research Council (ERC) under the European Union’s Horizon Europe Programme WePhICom (grant agreement number 101225119), as well as by the European Union through the Horizon Europe project COVER under grant agreement number 101086228, the Horizon Europe project UNITE under grant agreement number 101129618, the Horizon Europe project INSTINCT under grant agreement number 101139161, and the Horizon Europe project TWIN6G under grant agreement number 101182794, as well as by the Agence Nationale de la Recherche (ANR) through the France 2030 project ANR-PEPR Networks of the Future under grant agreement NF-Founds 22-PEFT-0010, and by the CHIST-ERA project PASSIONATE under grant agreements CHIST-ERA-22-WAI-04 and ANR-23-CHR4-0003-01. His work was also supported in part by the Engineering and Physical Sciences Research Council (EPSRC), part of UK Research and Innovation, and the UK Department of Science, Innovation and Technology through the CHEDDAR Telecom Hub under grant EP/Y037421/1, through the HASC Telecom Hub under grant EP/Y037197/1, and through the TITAN Telecom Hub under grant EP/Y037243/1.}
\vspace{-2mm}}

\begin{document}

\maketitle

\begin{abstract}
Holographic multiple-input multiple-output (MIMO) enables electrically large continuous apertures, overcoming the physical scaling limits of conventional MIMO architectures with half-wavelength spacing. Their near-field operating regime requires channel models that jointly capture line-of-sight (LoS) and non-line-of-sight (NLoS) components in a physically consistent manner. Existing studies typically treat these components separately or rely on environment-specific multipath models. In this work, we present a comprehensive analysis of the wavenumber-division multiplexing (WDM) framework, originally proposed for purely LoS channels, applied to a unified electromagnetic (EM)-based LoS+NLoS channel model. We first formulate the EM-based channel model for holographic lines and then apply WDM to this model, enabling a comparison between the corresponding spatial-sampling-based and expansion-based formulations. Applying WDM to the NLoS component yields its angular-domain representation, enabling direct characterization through the power spectral factor and power spectral density. We further derive closed-form characterizations for isotropic and non-isotropic scattering, with the former recovering Jakes’ isotropic model. Lastly, we evaluate the resulting degrees of freedom and ergodic capacity, showing that incorporating the NLoS component substantially improves the performance relative to the purely LoS case.
\end{abstract}


\section{Introduction}
\label{sec:Intro}

The rapid increase in wireless data demand necessitates a shift towards higher frequency bands and a fundamental redesign of transceiver architectures \cite{Atz25}. To achieve this growth, scaling traditional massive \ac{MIMO} systems by simply adding more antenna elements is impractical, as maintaining the required half-wavelength spacing to mitigate spatial correlation would demand prohibitively large arrays \cite{wei2024electromagnetic}. Holographic \ac{MIMO} addresses this limitation by enabling electrically large continuous apertures (i.e., surfaces or lines) or dense antenna arrays that allow a fine-grained control of the \ac{EM} waves \cite{pizzo2022spatial}. Related concepts include large intelligent surfaces \cite{hu2018beyond, decarli2021communication}, reconfigurable intelligent surfaces \cite{danufane2021path}, and holographic radio systems \cite{prather2017optically}. This paper considers ideal holographic \ac{MIMO} with continuous apertures, although practical implementations typically consist of nearly continuous apertures realized using densely packed metamaterial elements or \acp{TCA} \cite{gong2024near}. Such implementations are flexible, scalable, and energy efficient \cite{omam2025holographic}, but inevitably introduce mutual coupling effects \cite{gong2023holographic}. Moreover, they are subject to hardware limitations: for example, metamaterial-based architectures suffer from impairments such as substrate losses and signal leakage, while \ac{TCA}-based architectures are sensitive to environmental variations \cite{gong2023holographic}. The electrically large nature of holographic apertures also has important implications for propagation modeling by inherently shifting the propagation regime from the far field to the near field, where the classical plane-wave model \cite{sayeed2002deconstructing, poon2005degrees} becomes invalid. Nevertheless, spherical waves can still be expressed as an infinite superposition of plane waves \cite{hansen1999plane, chew1999waves}, enabling tractable modeling of near-field propagation.

\subsection{Related Work and Motivation}
\label{sec: Motivation and Related Work}

The existing holographic \ac{MIMO} literature predominantly treats the \ac{LoS} and \ac{NLoS} components separately. \ac{LoS} propagation has been examined from several viewpoints. For instance, \cite{pizzo2020holographic, pizzo2022spatial} formulated a point-to-point \ac{LoS} channel model using a plane-wave representation, providing an angular-domain description in terms of impulse functions. The work in \cite{jin2024achievable} studied a holographic line model and analyzed the achievable rate from an information-theoretic perspective. The study in \cite{tang2023line} compared the normalized singular values of the angular- and spatial-domain \ac{LoS} channels through ray-tracing, relying on unitary equivalence without introducing an \ac{EM}-based \ac{LoS} channel model. Furthermore, \cite{di2023mimo} considered both \ac{MIMO} arrays and holographic surfaces, comparing their spatial multiplexing capability in a \ac{LoS} scenario. In addition, \ac{WDM} was proposed in \cite{sanguinetti2022wavenumber,d2022performance} as the spatial-frequency counterpart of \ac{OFDM} for purely \ac{LoS} channels. While \ac{OFDM} enables multiplexing in the frequency domain by transmitting time-domain signals over orthogonal frequency subchannels, \ac{WDM} enables multiplexing in the wavenumber (spatial-frequency) domain through an orthogonal decomposition of the spatially continuous transmit current and received field. Fourier basis functions, although not necessarily optimal \cite{miller2000communicating}, were used for this purpose due to their analytical tractability and efficient hardware implementation at the \ac{EM} level using lens antenna arrays or metasurfaces \cite{li2015surface,tretyakov2015metasurfaces}. \ac{WDM} was extended to holographic surfaces in \cite{iacovelli2025}, still considering \ac{LoS} propagation.

On the other hand, modeling \ac{NLoS} propagation requires characterizing how the environment perturbs the transmitted \ac{EM} field. In principle, this demands solving Fredholm integral equations of the second kind \cite{chew1999waves}, which are analytically intractable. To obtain tractable models for field-medium interactions, Born-type approximations have been widely adopted \cite{pizzo2020holographic, pizzo2022fourier, pizzo2022spatial, pizzo2020spatially}, enabling stochastic \ac{NLoS} channel models for holographic surfaces under both isotropic and non-isotropic scattering, with the latter typically modeled as a mixture of \ac{3D} \ac{vMF} distributions. For holographic lines, isotropic scattering was considered in \cite{pizzo2020spatially}. The combined \ac{LoS}+\ac{NLoS} channel has also been examined. Specifically, \cite{zhang2025fundamental} relied on existing \ac{LoS} and \ac{NLoS} formulations from \cite{pizzo2020holographic} and \cite{pizzo2020spatially} to derive closed-form bounds and approximations for the ergodic mutual information and outage probability for holographic planar arrays using infinite-dimensional random matrix theory. Moreover, \cite{lu2023near, dardari2021nlos} presented deterministic ray-tracing-based models tied to the specific propagation environment.

As highlighted above, existing studies have largely focused on either purely \ac{LoS} or purely \ac{NLoS} channels, while works including multipath typically rely on deterministic, environment-specific models for modeling the \ac{LoS} and \ac{NLoS} components. In practice, millimeter-wave and sub-terahertz systems are often \ac{LoS}-dominated, yet small-scale fading remains relevant. Hence, \ac{NLoS} components cannot be entirely neglected, and their joint behavior with \ac{LoS} should be analyzed consistently. In this work, we study the \ac{WDM} framework in the \ac{LoS}+\ac{NLoS} scenario. Motivated by the seminal work on \ac{WDM} \cite{sanguinetti2022wavenumber}, which specifically focused on continuous holographic lines under \ac{LoS} propagation, we adopt the same linear geometry to analyze the impact of the combined \ac{LoS}+\ac{NLoS} channel. Since the aforementioned existing \ac{LoS} and \ac{NLoS} formulations cannot be directly applied to analyze the considered \ac{WDM} framework, we revisit and adapt the formulations in \cite{pizzo2022spatial,pizzo2022fourier} by deriving the holographic line model from the corresponding surface model, thereby providing a unified \ac{LoS}+\ac{NLoS} channel formulation for holographic lines. From a practical implementation perspective, the adoption of holographic lines instead of the more widely studied holographic surfaces provides a balanced trade-off between performance, complexity, and analytical tractability. Holographic surfaces enable full \ac{3D} control of \ac{EM} waves, supporting narrow pencil beams, flexible multi-beam patterns, and advanced wavefront shaping, but at the cost of significantly increased design, calibration, and signal-processing complexity. In contrast, holographic lines offer \ac{2D} control and generate fan-shaped beams \cite{boyle2002radiation}, making them well suited for low-complexity scanning and sensing applications while retaining cost and power efficiency. The choice between line and surface architectures ultimately depends on the target application, with hybrid solutions emerging to combine the flexibility of surfaces with the practicality of lines \cite{guo2025antenna}. From an analytical perspective, the line model captures the essential physical characteristics of holographic \ac{MIMO} systems and serves as a fundamental building block of holographic surfaces.

The \ac{MIMO} representation of continuous apertures is achieved either through spatial sampling of the transmit and receive apertures, resulting in a spatial-sampling-based formulation, or by expanding the spatially continuous transmit current and received field using suitable basis functions, resulting in an expansion-based formulation. In this work, we present a unified perspective combining both formulations. The expansion-based formulation naturally leads to the \ac{WDM} framework \cite{sanguinetti2022wavenumber}, and the holographic line model provides a particularly tractable setting for its analysis. This motivates our focus on holographic lines and the use of \ac{WDM} as an analytical tool for investigating holographic \ac{MIMO} channels.

\subsection{Contributions}
\label{sec: Contributions}

In this work, we present a comprehensive analysis of the \ac{WDM} framework applied to a unified \ac{EM}-based \ac{LoS}+\ac{NLoS} channel model for holographic lines. This analysis can be extended to holographic surfaces at the cost of more involved derivations and without altering the main analytical insights. Our contributions can be broadly divided into two parts: first, we develop the \ac{EM}-based \ac{LoS}+\ac{NLoS} channel model for holographic lines; then, based on this model, we extend the \ac{WDM} framework to the \ac{LoS}+\ac{NLoS} scenario, enabling a comparison between the corresponding spatial-sampling-based and expansion-based formulations. The \ac{EM}-based channel formulation builds on concepts from \cite{pizzo2022spatial,pizzo2022fourier}, while requiring nontrivial derivations and adaptation to the considered system model.

The main contributions are summarized as follows.
\begin{itemize}[leftmargin=5mm]
    \item We begin by demonstrating that the holographic surface model degenerates into its corresponding line model. Using this result, we provide a comprehensive characterization of the \ac{EM}-based \ac{LoS}+\ac{NLoS} channel for holographic lines, where the corresponding channel matrix follows from spatial sampling. In the process, we rigorously verify that the \ac{LoS} and \ac{NLoS} components combine additively even in the near field, an intuitive yet previously unverified property in holographic \ac{MIMO}. The \ac{LoS} component is analyzed using both ray-tracing-based and \ac{EM}-based models. The \ac{NLoS} component is characterized under both isotropic and non-isotropic scattering, with the latter modeled as a mixture of \ac{2D} \ac{vMF} distributions. We further derive closed-form expressions for the \ac{PSD} and the kernel of the \ac{LoS}+\ac{NLoS} channel.

    \item We extend the \ac{WDM} framework to the \ac{EM}-based \ac{LoS}+\ac{NLoS} channel model. For the purely \ac{LoS} channel, we show that the normalized eigenvalues obtained from the ray-tracing-based, \ac{EM}-based, and \ac{WDM}-applied models closely match and consistently yield the same \ac{DoF}. On the other hand, applying \ac{WDM} to the \ac{EM}-based \ac{NLoS} channel yields the corresponding angular-domain representation. Consequently, the spatial-sampling-based and expansion-based formulations yield equivalent eigenvalue spectra for both the \ac{LoS} and \ac{NLoS} components, differing only by a constant scaling. Moreover, the \ac{WDM}-applied \ac{NLoS} channel enables an analytical characterization of the relation between its \ac{PSD} and \ac{PSF}.

    \item We complete the analysis of the \ac{EM}-based and \ac{WDM}-applied \ac{NLoS} channels by characterizing the resulting angular-domain channel. Then, we derive closed-form expressions for the \ac{ACF} and \ac{PSD} under both isotropic and non-isotropic scattering, with the isotropic case recovering the classical Jakes' isotropic model. The results demonstrate that \ac{WDM} serves not only as a multiplexing technique but also as an effective analytical tool for studying holographic \ac{MIMO} channels.

    \item We define \ac{DoF} for the considered \ac{LoS}+\ac{NLoS} scenario and evaluate the ergodic capacity, showing that incorporating the \ac{NLoS} component provides substantial performance gains over the purely \ac{LoS} channel. Under isotropic scattering, the additional \ac{NLoS} component strengthens all the weaker eigenmodes of the \ac{LoS} channel, leading to a more balanced eigenvalue distribution and thus higher capacity. In contrast, non-isotropic scattering concentrates power in the directions around the scattering clusters, reinforcing a limited subset of eigenmodes and resulting in noticeably lower capacity than in the isotropic case.
\end{itemize}

Part of this work was presented in our conference paper \cite{Praj25}, which analyzed \ac{WDM} with purely \ac{NLoS} channels.

\smallskip

\textbf{\textit{Outline.}} The rest of the paper is organized as follows. Section~\ref{sec: sys} defines the system model. Section~\ref{sec:holographicMIMO} develops the \ac{EM}-based \ac{LoS}+\ac{NLoS} channel model for holographic lines. Section~\ref{sec: Wavenumber-Division Multiplexing} analyzes the \ac{WDM}-applied scenario. Section~\ref{sec: Calculation of Channel Coefficients for WDM and holographic channel} completes the previous analysis by characterizing the angular-domain \ac{NLoS} channel. Lastly, Section~\ref{sec: Numerical Results} presents the numerical results, whereas Section~\ref{sec: Conclusions} concludes the paper and outlines possible future directions.

\smallskip

\textbf{\textit{Notation.}} Boldface lowercase and uppercase letters denote vectors and matrices, respectively, whereas calligraphic letters represent sets. The conjugate, transpose, and Hermitian transpose operators are denoted by $(\cdot)^{*}$, $(\cdot)^{\tran}$, and $(\cdot)^{\herm}$, respectively. $[\A] _{u,v}$ represents the $(u, v)$-th entry of $\A$. The Euclidean norm and vectorization operator are represented by $\|\cdot\|$ and $\vec(\cdot)$, respectively. The Hadamard (entry-wise) and Kronecker products of two matrices are denoted by $\odot$ and $\otimes$, respectively. The determinant and square root of a square matrix are represented by $\det(\cdot)$ and $(\cdot)^{\frac{1}{2}}$, respectively. The $n$-dimensional identity matrix is denoted by $\I_{n}$. $\diag(\cdot)$ (resp. $e^{j\diag(\cdot)}$) produces a diagonal matrix with the vector argument (resp. the complex exponential of the vector argument) on its diagonal. The set of real positive numbers is denoted by $\Real_+$. The Cartesian product of two sets is denoted by $\Cross$. The indicator function for a set $\mathcal{D}$ is defined as $\mathbbm{1}_{\mathcal{D}}(x)$, which is equal to $1$ if $x \in \mathcal{D}$ and to $0$ otherwise. The cardinality of a set is represented by $\card(\cdot)$. $\setN_{\mathbb{C}}(\0,\A)$ represents the circularly symmetric complex Gaussian distribution with zero mean and covariance matrix $\A$. The mathematical expectation operator is denoted by $\mathbb{E}[ \cdot ]$. $j = \sqrt{-1}$ is the imaginary unit. The absolute value and floor function are represented by $|\cdot|$ and $\left\lfloor\cdot\right\rfloor$, respectively. The sinc function is denoted by $\mathrm{sinc}(x)=\frac{\mathrm{sin}(\pi x)}{\pi x}$. The Dirac and Kronecker delta functions are represented by $\delta(\cdot)$ and $\delta[ \cdot ]$, respectively. The time index is denoted by $t$ and the angular frequency is represented by $\omega$.

\section{System Model}
\label{sec: sys}

Consider a point-to-point holographic \ac{MIMO} system as depicted in Fig.~ \ref{fig:Fig1}, where a line source spanning the linear region $\setL_{\textrm{s}} \subset \Real^{2}$ with length $L_{\textrm{s}}$ transmits data to a line receiver spanning the linear region $\setL_{\textrm{r}} \subset \Real^{2}$ with length $L_{\textrm{r}}$. The two lines are parallel and oriented along the $x$-axis, with their centers aligned along the $z$-axis and separated by a distance $d$. Let $\s = [s_{x}, s_{z}]^{\tran} \in \setL_{\textrm{s}}$ and $\r = [r_{x}, r_{z}]^{\tran} \in \setL_{\textrm{r}}$ denote arbitrary points within the source and receiver regions, respectively. Throughout the paper, we assume that the communication takes place via scalar waves as in, e.g., \cite{pizzo2020holographic,pizzo2020degrees,pizzo2022fourier,pizzo2020spatially,pizzo2022spatial,decarli2021communication,an2023tutorial,wei2022multi}. This assumption simplifies the analysis by allowing the use of the scalar Green's function; in this regard, Appendix~\ref{sec: Different type of Green functions and their approximation} provides an empirical justification based on a comparison of amplitudes profiles obtained with different Green's functions. For notational convenience, we introduce the transmit and receive wavenumbers $\kappa$ and $k$, respectively, defined as $\kappa = k = \frac{2\pi}{\lambda}$ and measured in radians per meter, where $\lambda$ denotes the wavelength. The wave impedance is given by $\eta = \sqrt{\mu / \epsilon}$, where $\mu$ and $\epsilon$ are the permeability and permittivity of the medium, respectively (e.g., we have $\eta \simeq 120\pi$~ohms in free space) \cite{balanis2016antenna}.

Assume $N_{\textrm{s}}$ and $N_{\textrm{r}}$ uniformly distributed sampling points within the regions $\setL_{\textrm{s}}$ and $\setL_{\textrm{r}}$, respectively, following the Nyquist spatial sampling criterion, with spatial sampling spacings $\Delta_{\textrm{s}}$ and $\Delta_{\textrm{r}}$ at the source and receiver, respectively. Let $\s_v = [s_{x_v}, s_{z_v}]^{\tran} \in \Real^2$, for $v = 1, \ldots, N_{\textrm{s}}$, and $\r_u = [r_{x_u}, r_{z_u}]^{\tran} \in \Real^2$, for $u = 1, \ldots, N_{\textrm{r}}$, denote the coordinates of the $v$-th and $u$-th sampling points at the source and receiver, respectively.
\begin{figure}[t]
     \centering
     \begin{tikzpicture}[>=latex]

\small

\def\dh{1cm} 
\def\dv{1cm} 
\node[] (center) at (0,0) {}; 

\draw[->, thick, red] (-4.25,-0.75) -- (-4.25,0) node[midway, sloped, above, xshift=2pt] {\footnotesize $x$-axis};
\draw[->, thick, red] (-4.25,-0.75) -- (-3.5,-0.75) node[midway, sloped, below, xshift=2pt] {\footnotesize $z$-axis};
\fill[red] (-4.25,-0.75) circle (1pt);
\draw[red, thick] (-4.25,-0.75) circle (3pt);
\node[rotate=45, anchor=east] at (-4.3,-0.8) {\footnotesize \textcolor{red}{$y$-axis}};

\draw[very thick] (-2.50*\dh,-1.5*\dv) -- (-2.50*\dh,1.5*\dv) node[midway, sloped, below, yshift=-2pt] {\footnotesize line source};
\foreach \y in {-1.5,-1,...,1.5} {
    \node at (-2.50*\dh,\y*\dv) {$\times$};
};

\draw[very thick] (2.50*\dh,-1.5*\dv)--(2.50*\dh,1.5*\dv) node[midway, sloped, above, yshift=2pt] {\footnotesize line receiver};
\foreach \y in {-1.5,-1,...,1.5} {
    \node at (2.50*\dh,\y*\dv) {$\times$};
}

\node[draw, minimum width=0.8*\dh, minimum height=0.5*\dv, fill=gray!20, ellipse, rotate=60, transform shape] () at (0,1*\dv) {};

\node[draw, minimum width=0.5*\dh, minimum height=0.4*\dv, fill=gray!20, ellipse, rotate=120, transform shape] () at (0,-1*\dv) {};

\draw[<->, black] (-3*\dh,-1.5*\dv) -- (-3*\dh,1.5*\dv) node[midway, anchor=east] {$L_{\textrm{s}}$};

\draw[<->, black] (3*\dh,-1.5*\dv) -- (3*\dh,1.5*\dv) node[midway, anchor=west] {$L_{\textrm{r}}$};

\draw[<->, black] (-2.50*\dh,-1.8) -- (2.50*\dh,-1.8) node[midway, anchor=south] {$d$};

\draw[<->, black] (-2.25*\dh,1*\dv) -- (-2.25\dh,1.5*\dv) node[midway, anchor=west] {$\Delta_{\textrm{s}}$};

\draw[<->, black] (2.25*\dh,1*\dv) -- (2.25*\dh,1.5*\dv) node[midway, anchor=east] {$\Delta_{\textrm{r}}$};

\end{tikzpicture}
     \caption{A schematic of the considered \ac{LoS}+\ac{NLoS} holographic \ac{MIMO} system model. The crosses indicate the spatially sampled points.}
     \label{fig:Fig1}
 \end{figure}
At any symbol time, the system model in Fig.~\ref{fig:Fig1} can be described by the equivalent discrete space model \cite{tse2004fundamentals}, with received signal given by
\begin{align}
    \label{eqn: received signal}
    \y= \H \x+ \z \in \mathbb{C}^{N_{\textrm{r}}},
\end{align}
where $\H\in \mathbb{C}^{N_{\textrm{r}}\times N_{\textrm{s}}}$ is the channel matrix, $ \x\in \mathbb{C}^{N_{\textrm{s}}}$ is the transmitted signal, and $\z \sim \setN_{\mathbb{C}}(\0, \chi^{2} \I_{N_{\textrm{r}}})$ is a vector of \ac{AWGN} with variance $\chi^{2}$. We consider a general \ac{LoS}+\ac{NLoS} channel and introduce the \ac{LoS} and \ac{NLoS} channel matrices $\H^{\los}\in \mathbb{C}^{N_{\textrm{r}}\times N_{\textrm{s}}}$ and $\H^{\nlos}\in \mathbb{C}^{N_{\textrm{r}}\times N_{\textrm{s}}}$, respectively. The channel matrix admits the stochastic decomposition $\H=\H^{\los}+\H^{\nlos}$, as we rigorously verify in Section~\ref{sec: LoS+NLoS Channel Model}. Nevertheless, the particular form of $\H$ depends on the environment between the source and receiver, and one of the components may be absent.

\section{EM-Based Channel Model for Holographic Lines}
\label{sec:holographicMIMO}

In this section, we first derive the holographic line model introduced in Section~\ref{sec: sys} as a degenerate case of the corresponding surface model; see Section~\ref{sec:line}. Then, building on concepts from \mbox{\cite{pizzo2022spatial, pizzo2022fourier}}, we present a comprehensive characterization of the \ac{EM}-based \ac{LoS}+\ac{NLoS} channel for holographic lines, which is used to investigate the \ac{WDM}-applied scenario in Section~\ref{sec: Wavenumber-Division Multiplexing}. Specifically, we begin by considering a purely \ac{LoS} (i.e., deterministic) channel model in Section~\ref{sec: Line of Sight Channel Model}, followed by a purely \ac{NLoS} (i.e., stochastic) channel model in Section~\ref{sec: NLoS}. Once both the \ac{LoS} and \ac{NLoS} components are well understood individually, we study their combined behavior in Section~\ref{sec: LoS+NLoS Channel Model}.

\subsection{From Holographic Surfaces to Lines}
\label{sec:line}

In this section, we demonstrate how the holographic surface model degenerates into its corresponding line model building on the formulation in \cite{pizzo2022spatial, pizzo2022fourier}.

We consider a surface source and receiver spanning the planar regions $\setS_{\textrm{s}} \subset \Real^{3}$ and $\setS_{\textrm{r}} \subset \Real^{3}$, respectively. Let $\tilde{\s}=[s_x, s_y, s_z]^{\tran} \in \setS_{\textrm{s}}$ and $\tilde{\r}=[r_x,r_y,r_z]^{\tran} \in \setS_{\textrm{r}}$ denote arbitrary points within the source and receiver regions, respectively. For simplicity, and without loss of generality, we assume that the source point is located at the origin, i.e., $\tilde{\s}=[0,0,0]^{\tran}$. The (spatial-domain) channel impulse response of a surface source, as presented in \cite{pizzo2022spatial, pizzo2022fourier}, follows from solving the \ac{3D} Helmholtz wave equation \cite[Eq.~(2.2.19)]{chew1999waves}, which yields the scalar Green's function $\frac{e^{j\kappa \|\tilde{\tilde{\r}}\|}}{4\pi \|\tilde{\tilde{\r}}\|}$, with $\|\tilde{\r}\|=\sqrt{r_x^2+r_y^2+r_z^2}$. To obtain the line model as a degenerate case of the surface model, we start from Weyl's identity considering only the forward-traveling wave\footnote{Throughout the paper, we consider only forward-traveling waves, following a commonly adopted assumption in the literature \cite{pizzo2020holographic,pizzo2020spatially,pizzo2022spatial,pizzo2022fourier,zhang2025fundamental,lu2023near}. Evanescent waves are mainly relevant in the vicinity of the source and can be neglected at practical communication distances, even in the near field \cite{ouyang2024impact}, because their associated power decays exponentially while the corresponding electric and/or magnetic fields exhibit highly oscillatory behavior \cite{chew1999waves}. Among traveling waves, we are only interested in the forward propagation towards the receiver.}\label{fn:traveling_waves} (i.e., $r_z > 0$) and demonstrate that the planar region effectively collapses to a linear region. Hence, we have \cite[Eq.~(2.2.27)]{chew1999waves}
\begin{align}
\label{eqn: Weyl's identity}
\frac{e^{j\kappa \|\tilde{\r}\|}}{ \|\tilde{\r}\|}=\frac{j}{2\pi}\int_{\Real^2} \frac{e^{j\tilde{\kappab}^\tran \tilde{\r}}}{\tilde{\gamma}(\kappa_x, \kappa_y)} \diff \kappa_x \diff \kappa_y,
\end{align}
with $\tilde{\kappab}=[\kappa_x, \kappa_y, \tilde{\gamma}(\kappa_x, \kappa_y)]^\tran$ and $\tilde{\gamma}(\kappa_x, \kappa_y)=\sqrt{\kappa^2-\kappa_x^2-\kappa_y^2}$. As we are interested in the line model (see Fig.~\ref{fig:Fig1}), it is natural to integrate out $r_y$ from \eqref{eqn: Weyl's identity}, leading to
\begin{align}
\label{eqn: integration over y}
    \int_{-\infty}^\infty
    \frac{e^{jk\|\tilde{\r}\|}}{\|\tilde{\r}\|} \diff r_y
    =\frac{j}{2\pi}
    \int_{\Real^3} \frac{e^{j\tilde{\kappab}^\tran \tilde{\r}}}{\tilde{\gamma}(k_x, k_y)} \diff k_x \diff k_y \diff r_y.
\end{align}
Changing the order of integration (due to Fubini's theorem) and performing some algebraic manipulations, we obtain
\begin{align}
    \int_{-\infty}^\infty
    \frac{e^{j\kappa\|\tilde{\r}\|}}{\|\tilde{\r}\|} \diff r_y
    =\,&\frac{j}{2\pi}
    \int_{\Real^2} \frac{e^{j(\kappa_x r_x+\tilde{\gamma}(\kappa_x, \kappa_y)r_z)}}{\tilde{\gamma}(\kappa_x, \kappa_y)} \diff \kappa_x \diff \kappa_y \nonumber\\ 
    & \times\int_{-\infty}^\infty e^{j \kappa_y r_y} \diff r_y.
\end{align}
Noting that $\int_{-\infty}^\infty e^{j\kappa_y r_y} \diff r_y=2\pi \delta(\kappa_y)$, we rewrite the above triple integral as
\begin{align}
    \int_{-\infty}^\infty 
    \frac{e^{j\kappa\|\tilde{\r}\|}}{\|\tilde{\r}\|} \diff r_y
     =j\int_{\Real^2} \frac{e^{j(\kappa_x r_x+\tilde{\gamma}(\kappa_x, \kappa_y)r_z)}}{\tilde{\gamma}(\kappa_x, \kappa_y)} \delta(\kappa_y) \diff \kappa_x \diff \kappa_y,
\end{align}
which degenerates into a single integral as
\begin{align}
\label{eqn: integeral 2D}
    &\int_{-\infty}^\infty 
    \frac{e^{j\kappa\|\tilde{\r}\|}}{\|\tilde{\r}\|} \diff r_y =j
    \int_{-\infty}^\infty \frac{e^{j(\kappa_x r_x+\tilde{\gamma}(\kappa_x,0)r_z)}}{\tilde{\gamma}(\kappa_x,0)} \diff \kappa_x. 
\end{align}
Defining $\rho = \sqrt{r_x^{2} + r_z^{2}}$ and applying the change of variable $y = \rho \sinh{w}$ allows to simplify \eqref{eqn: integeral 2D} as
\begin{align}
    \label{eqn:simplified left}
    \int_{-\infty}^\infty 
    \frac{e^{j\kappa\|\tilde{\r}\|}}{\|\tilde{\r}\|} \diff r_y=\int_{-\infty}^{\infty} e^{j\kappa\rho\cosh{w}}\diff w.
\end{align}
Now, using the relation in \cite[Eq.~(8.421.1)]{edition2007table}, we obtain
\begin{align}
\label{eqn: Hankel equivalent to surface}
\int_{-\infty}^{\infty} e^{j\kappa\rho\cosh{w}}\diff w=j\pi H_0^{(1)}\left(\kappa\rho\right),
\end{align}
which, along with \eqref{eqn: integeral 2D}--\eqref{eqn:simplified left}, leads to
\begin{align}
\label{eqn: Hankel expansion}
 H_0^{(1)}\left(\kappa \rho\right)=  \frac{1}{\pi} 
 \int_{-\infty}^\infty \frac{e^{j(\kappa_x r_x+\tilde{\gamma}(\kappa_x,0)r_z)}}{\tilde{\gamma}(\kappa_x,0)} \diff \kappa_x,
\end{align}
where $H_{0}^{(1)}(\cdot)$ is the Hankel function of the first kind and order zero. Finally, multiplying both sides of \eqref{eqn: Hankel expansion} by $\frac{j}{4}$, the left-hand side becomes the solution of the \ac{2D} Helmholtz wave equation \cite[Eq.~(2.2.4)]{chew1999waves}, which is also recognized as the \ac{2D} scalar Green's function given by
\begin{align}
\label{eqn: Green's function for a line source}
    g(\rho)=\frac{j}{4}H_0^{(1)}(\kappa \rho).
\end{align}
Moreover, the right-hand side of \eqref{eqn: Hankel expansion}, after multiplication by $\frac{j}{4}$, represents the spectral decomposition of \eqref{eqn: Green's function for a line source}, also referred to as the plane-wave representation \cite[Eq.~(2.2.10)]{chew1999waves}. The formulation in \eqref{eqn: Hankel expansion} and \eqref{eqn: Green's function for a line source} can be alternatively derived directly from \cite[Eq.~(2.2.1)]{chew1999waves}.

From \eqref{eqn: Green's function for a line source}, it can be observed that the planar region $\setS_{\textrm{r}}$ collapses to the linear region $\setL_{\textrm{r}}$; similarly, $\setS_{\textrm{s}}$ would collapse to $\setL_{\textrm{s}}$ if the considered source point was not assumed to be located at the origin. Hence, from an \ac{EM} wave perspective, a line source can be interpreted as a degenerate case of a surface source. Consequently, to derive the line counterparts of the results in \cite{pizzo2022spatial}, it is sufficient to replace $\diff k_y$ with $2\pi \delta(k_y) \diff k_y $ in all the derivations pertaining to the surface model.

\subsection{LoS Channel Model}
\label{sec: Line of Sight Channel Model}

In this section, we present the \ac{EM}-based \ac{LoS} channel model for holographic lines. This model builds on concepts from \cite{pizzo2020holographic}, which we adapt to our system configuration and extend by deriving a new closed-form expression for the channel. We begin by introducing the well-established ray-tracing-based model (see, e.g., \cite{do2022line,do2020reconfigurable}) for comparison. In the setup illustrated in Fig.~\ref{fig:Fig1}, the distance between two sampling points $v$ and $u$ at the source and receiver, respectively, is given by
\begin{align}
\label{eqn: Ray Tracing}
r_{u,v}= \sqrt{d^2+(u\Delta_{\textrm{r}}-v\Delta_{\textrm{s}})^2}.
\end{align}
Hence, the corresponding complex channel gain, which gives rise to spherical wavefronts in the near field, is given by \cite{song2015spatial}
\begin{align}
\label{eqn:Ray Tracing Channel Gain}
h^{\los}_{u, v}=\big[\H^{\los}\big]_{u, v}= \frac{\lambda}{4\pi r_{u,v}}e^{j\kappa r_{u,v}}. 
\end{align}
Building on the framework presented in \cite{pizzo2020holographic} and extending it to account for the degeneration of a surface into a line, the \ac{LoS} channel impulse response can be expressed as
\begin{align}
\label{eqn: LoS}
h^{\los}(\r, \s) = -j\kappa\eta g\big(\|\r-\s\|\big),
\end{align}
where $g\big(\|\r-\s\|\big)$ is the scalar Green's function between the source and receive points $\s$ and $\r$, with distance $\|\r-\s\|$ (cf. \eqref{eqn: Green's function for a line source}). Plugging \eqref{eqn: Green's function for a line source} into \eqref{eqn: LoS} yields the closed-form expression
\begin{align}
    \label{eqn:Closed LoS}
h^{\los}(\r, \s)=\frac{\kappa\eta}{4}H_{0}^{(1)}(\kappa\|\r-\s\|).
\end{align}
Now, considering holographic lines sampled as described in Section~\ref{sec: sys}, the complex channel gain $h^{\los}_{u,v}$ is obtained as
 \begin{align}
    \label{eqn:Closed LoS Matrix}
h^{\los}_{u, v}=\frac{\kappa\eta}{4}H_{0}^{(1)}(\kappa r_{u, v}),
\end{align}
with $r_{u, v}$ defined in \eqref{eqn: Ray Tracing}. Based on \eqref{eqn: Hankel expansion} and \eqref{eqn:Closed LoS}, the channel impulse response admits the form
\begin{align}
    \label{eqn: LoS Channel}
h^{\los}(\r, \s)=\frac{\kappa\eta}{4\pi} \int_{-\infty}^{\infty}\frac{e^{j\kappab^{\tran}(\r-\s)}}{\gamma(\kappa_x)}\diff\kappa_x, \ r_z>s_z,
\end{align}
where $\kappab=[\kappa_{x},\gamma(\kappa_{x})]^{\tran} \in \Real^2$ is the wave vector corresponding to the transmit propagation direction $\hat{\kappab} =\frac{\kappab}{\|\kappab\|}$ and the function $\gamma(\kappa_{x})$ is defined as
\begin{align}
\label{eqn:gamma}
    \gamma(\kappa_{x})=
    \begin{cases}
      \sqrt{\kappa^{2}-\kappa_{x}^{2}}, & \kappa_{x}\in \setD \\
      j\sqrt{\kappa^{2}-\kappa_{x}^{2}}, & \kappa_{x}\in \setD^{\textrm{c}},
    \end{cases}
\end{align}
which is real-valued within the support
\begin{align}
\label{eqn:support}
\setD=\{\kappa_{x}\in \Real: -\kappa\leq\kappa_{x}\leq \kappa\}
\end{align}
and imaginary-valued on the complementary set $\setD^{\textrm{c}}$. We note that $\setD$ and $\setD^{\textrm{c}}$ correspond to traveling and evanescent waves, respectively.

Inspection of \eqref{eqn: LoS Channel} reveals that a point source at $\s$ radiates an infinite number of plane waves. Let $\k=[k_{x}, \gamma(k_{x})]^{\tran} \in \Real^2$ be the wave vector corresponding to the receive propagation direction $\hat{\k} =\frac{\k}{\|\k\|}$. In a purely \ac{LoS} scenario, where no scattering objects are present, there exists a one-to-one correspondence between the transmit and receive propagation directions, i.e., $\hat{\kappab}=\hat{\k}$. Therefore, \eqref{eqn: LoS Channel} can be expressed using the \ac{2D} plane-wave representation as \cite{pizzo2020spatially} 
\begin{align}
\label{eqn: LoS Plane Wave}
h^{\los}(\r,\s)=\frac{1}{2\pi}\int_{\Real^2}a_{\textrm{r}}(\k,\r)H_{\textrm{a}}(k_{x}, \kappa_{x})a_{\textrm{s}}(\kappab,\s)\diff k_{x}\diff\kappa_{x},
\end{align}
where
\begin{align}
\label{eqn: Angular LoS}
H_{\textrm{a}}(k_{x}, \kappa_{x})=\frac{\kappa\eta}{2}\frac{\delta(k_{x}-\kappa_{x})}{\gamma(\kappa_{x})}
\end{align}
represents the angular-domain channel impulse response and
\begin{subequations}
\label{eqn:plane waves source and receiver}
\begin{align}
a_{\textrm{s}}(\kappab,\s) & =e^{-j\kappab^{\tran}\s}=e^{-j(\kappa_{x}s_{x}+\gamma(\kappa_{x})s_{z})}, \\
a_{\textrm{r}}(\k,\r) & =e^{j\k^{\tran}\r}=e^{j(k_{x}r_{x}+\gamma(k_{x})r_{z})}
\end{align}
\end{subequations}
are the transmit and receive plane waves, respectively.

Having characterized the \ac{LoS} channel through the \ac{2D} plane-wave representation in \eqref{eqn: LoS Plane Wave}, we now adapt $H_{\mathrm{a}}(k_{x}, \kappa_{x})$ to account for the \ac{NLoS} component and study the \ac{NLoS} channel.

\subsection{NLoS Channel Model}
\label{sec: NLoS}

In this section, we present the \ac{EM}-based \ac{NLoS} channel model for holographic lines. Although the line model can be obtained as a degenerate case of the corresponding surface model, as shown in Section~\ref{sec:line}, the \ac{NLoS} channel for holographic lines has not been investigated in the existing literature. To characterize the \ac{NLoS} propagation environment, we employ the \ac{2D} \ac{vMF} distribution function, which is discussed in detail in Section~\ref{sec: Calculation of Channel Coefficients for WDM and holographic channel}.

In the presence of scatterers between the holographic lines, the one-to-one correspondence between the plane waves at the source and receiver no longer holds. To circumvent this difficulty, \eqref{eqn: LoS Plane Wave} is generalized using the propagation kernel $K(k_{x}, \kappa_{x})$ \cite[Eq.~(38)]{pizzo2022spatial}, which accounts for the interactions between the waves and the scatterers. Consequently, following the formulation in \cite{pizzo2022spatial}, the corresponding angular-domain channel impulse response is defined as
\begin{align}
    \label{eqn:General Ha}
H_{\textrm{a}}(k_{x}, \kappa_{x})=\frac{\kappa\eta}{2}\frac{K(k_{x},\kappa_{x})}{\sqrt{\gamma(k_{x})\gamma(\kappa_{x})}},
\end{align}
with $(k_{x}, \kappa_{x}) \in \setD^{2}$ and $\setD^{2} = \setD\Cross\setD$. Note that, for $K(k_{x}, \kappa_{x}) = \delta(k_x - \kappa_x)$, \eqref{eqn:General Ha} reduces to the angular-domain channel impulse response corresponding to $h^{\los}(\r, \s)$ (cf. \eqref{eqn: Angular LoS}).

Building on \eqref{eqn:General Ha}, we now model the \ac{NLoS} channel impulse response $h^{\nlos}(\r, \s)$ by accounting for the random interactions of the \ac{EM} waves with the scatterers through a stochastic characterization of the propagation environment. Specifically, we start from the \ac{2D} plane-wave representation in \eqref{eqn: LoS Plane Wave}. Then, we characterize $h^{\nlos}(\r,\s)$ through the stochastic modeling of $K(k_{x}, \kappa_{x})$ and derive an expression for the \ac{PSD} of $h^{\nlos}(\r,\s)$, which is then used to fully define the \ac{NLoS} channel. Finally, we apply the Fourier plane-wave series expansion and construct the corresponding \ac{NLoS} channel matrix.

\smallskip

\textbf{\textit{Stochastic characterization.}} The propagation kernel $K(k_{x}, \kappa_{x})$, which generates a spatially stationary circularly symmetric complex Gaussian random field $h^{\nlos}(\r,\s)$, can be written as \cite{pizzo2022spatial}
\begin{align}
    \label{eqn:NLoS_Kernel}
    K(k_{x}, \kappa_{x}) = A(k_{x}, \kappa_{x}) W(k_{x}, \kappa_{x}),
\end{align}
where $A(k_{x}, \kappa_{x})$ is a non-negative function characterizing the scattering environment, referred to as the \ac{PSF}, and $W(k_{x}, \kappa_{x})$ denotes the spatially stationary complex white Gaussian noise random field. In particular, the \ac{PSF} characterizes the coupling strength between the source and receiver, and is defined in $\setD^{2}$. Now, in view of \eqref{eqn:General Ha} and \eqref{eqn:NLoS_Kernel}, the corresponding angular-domain channel impulse response takes the form\footnote{Combining \eqref{eqn:plane waves source and receiver}, \eqref{eqn:AngularChannelNLoS}, and \eqref{eqn:NLoSSpatial}, the wavenumber-domain channel can be expressed as $e^{j\gamma(k_{x})r_{z}} H_{\textrm{a}}(k_{x}, \kappa_{x}) e^{-j\gamma(\kappa_{x})s_{z}}$, which maps the wavenumber (spatial-frequency) domain to the spatial domain through Fourier harmonics.\label{fn:ang_wave}}
\begin{align}
\label{eqn:AngularChannelNLoS}
    H_{\textrm{a}}(k_{x}, \kappa_{x}) = \frac{\kappa\eta}{2}\frac{A(k_{x}, \kappa_{x}) W(k_{x}, \kappa_{x})}{\sqrt{\gamma(k_{x}) \gamma(\kappa_{x})}}.
\end{align}
Therefore, plugging \eqref{eqn:AngularChannelNLoS} into the structure in \eqref{eqn: LoS Plane Wave}, the \ac{NLoS} channel impulse response can be written as
\begin{align}
\label{eqn:NLoSSpatial}
h^{\nlos}(\r,\s)= \, & \nonumber\frac{1}{2\pi}\int_{\setD^{2}}a_{\textrm{r}}(\k,\r)\frac{A(k_{x}, \kappa_{x})W(k_{x}, \kappa_{x})}{\sqrt{\gamma(k_{x}) \gamma(\kappa_{x})}}\\
& \times a_{\textrm{s}}(\kappab,\s)\diff k_{x}\diff\kappa_{x},
\end{align}
with the constant $\frac{\kappa\eta}{2}$ now absorbed in $A(k_{x}, \kappa_{x})$ for simplicity. Following a similar approach as in \cite{pizzo2022spatial}, \eqref{eqn:NLoSSpatial} can be expressed with a slight abuse of notation as
\begin{align}
\label{eqn:NLoSSpatialRandomProcess}
h^{\nlos}(\r,\s)= \, & \nonumber\frac{1}{2\pi}\int_{\Real^2}e^{jk_{x} r_x}
\sqrt{S(k_{x},\kappa_{x})} W(k_{x},\kappa_{x})
\\& \times e^{-j\kappa_{x}s_x} \diff k_{x}\diff\kappa_{x},
\end{align}
where \eqref{eqn:NLoSSpatial} and \eqref{eqn:NLoSSpatialRandomProcess} are statistically equivalent and 
\begin{align}
    \label{eqn:PSD}
    S(k_{x}, \kappa_{x})=\frac{A^2(k_{x}, \kappa_{x})}{\gamma(k_{x}) \gamma(\kappa_{x})}\mathbbm{1}_{\setD^2}(k_{x}, \kappa_{x})
\end{align}
is the \ac{PSD} of $h^{\nlos}(\r,\s)$. Consequently, we have 
\begin{align}
\label{eqn: total power}
\mathbb{E}\big[|h^{\nlos}(\r, \s)|^2\big]=\frac{1}{(2\pi)^2}\int_{\setD^2}S(k_{x}, \kappa_{x})\diff k_x \diff \kappa_x,
\end{align}
which represents the total power of $h^{\nlos}(\r, \s)$. In Section~\ref{sec: Wavenumber-Division Multiplexing}, we employ \eqref{eqn:NLoSSpatial} to derive the \ac{PSD} for the \ac{WDM}-applied \ac{NLoS} channel, which is consistent with \eqref{eqn:PSD}. Next, we apply the Fourier plane-wave series expansion approach from \cite{pizzo2020spatially} to express \eqref{eqn:NLoSSpatialRandomProcess} as a discrete sum rather than an integral.

\smallskip

\begin{figure*}
\setcounter{equation}{38}
\begin{align}
\label{eqn:LoS NLoS Spatial Random Process}
h(\r,\s)=\frac{1}{2\pi}\int_{\Real^2}\tilde{\mu} a_{\textrm{r}}(\k,\r)\delta(k_{x}-\kappa_{x})a_{\textrm{s}}(\kappab,\s)\diff k_{x}\diff\kappa_{x}+\frac{1}{2\pi}\int_{\Real^2}a_{\textrm{r}}(\k,\r)\sqrt{S(k_{x}, \kappa_{x})}W(k_{x}, \kappa_{x})a_{\textrm{s}}(\kappab,\s)\diff k_{x}\diff\kappa_{x}
\end{align}
\setcounter{equation}{28}
\hrule \vspace{-4mm}
\end{figure*}

\textbf{\textit{Fourier plane-wave series expansion.}} The Fourier plane-wave series expansion provides a discrete representation of a band-limited random process. Since the \ac{NLoS} channel in this work is modeled as a spatially stationary complex Gaussian random field, we use this representation to express $h^{\nlos}(\r,\s)$ in discrete form. Since $h^{\nlos}(\r,\s)$ is band-limited within $\setD^{2}$, sampling this support as in \cite{pizzo2020spatially} yields its discretized counterpart formed by the sets
\begin{subequations}
\label{eqn: transmit set and receive set}
\begin{align}
\setE_{\textrm{s}}& =\left\{p_x \in \mathbb{Z}:-\kappa \leq \frac{2\pi}{L_{\textrm{s}}}p_x\leq \kappa\right\},\\
\setE_{\textrm{r}}& =\left\{q_x \in \mathbb{Z}:-k \leq \frac{2\pi}{L_{\textrm{r}}}q_x\leq k \right\}
\end{align}
\end{subequations}
at the source and receiver, respectively. Now, define $ n_{\textrm{s}} = \card(\setE_{\textrm{s}}) =\big\lfloor\frac{2 L_{\textrm{s}}}{\lambda}\big\rfloor$ and $n_{\textrm{r}} = \card(\setE_{\textrm{r}}) =\big\lfloor\frac{2 L_{\textrm{r}}}{\lambda}\big\rfloor$ \cite{pizzo2020degrees}. Following the Fourier plane-wave series expansion of a spatially stationary complex Gaussian random process from \cite{pizzo2022fourier}, we have
\begin{align}
\label{eqn: Discrete NLoS}
h^{\nlos}(\r,\s)\approx\sum_{q_x\in\setE_{\textrm{r}}}\sum_{p_x \in\setE_{\textrm{s}}} H_{\textrm{a}}({q_x, p_x})a_{\textrm{r}}(q_x, \r)a_{\textrm{s}}(p_x, \s),
\end{align}
where
\begin{subequations}
\label{eqn: Discretize Plane wave source and receiver}
\begin{align}
a_{\textrm{s}}(p_x,\s)&=e^{-j\left(\frac{2\pi}{L_{\textrm{s}}}p_x s_{x}+\gamma(p_x)s_{z}\right)},\\
a_{\textrm{r}}(q_x,\r)&=e^{j\left(\frac{2\pi}{L_{\textrm{r}}}q_xr_{x}+\gamma(q_x)r_{z}\right)}
\end{align}
\end{subequations}
are the discretized counterparts of \eqref{eqn:plane waves source and receiver}, with $\gamma(p_x)=\sqrt{\kappa^{2}-\big(\frac{2\pi p_x}{L_{\textrm{s}}}\big)^2}$ and $\gamma(q_x)=\sqrt{k^{2}-\big(\frac{2\pi q_x}{L_{\textrm{r}}}\big)^2}$, and
\begin{align}
    \label{eqn: Discrete Angular}
    H_{\textrm{a}}(q_x, p_x)\sim \setN_{\mathbb{C}}(0,\sigma^2(q_x, p_x))
\end{align}
acts as a coupling coefficient characterizing the interaction between the transmit and receive plane waves. A detailed characterization of the variance $\sigma^2(q_x, p_x)$ is provided in Section~\ref{sec: Calculation of Channel Coefficients for WDM and holographic channel}.
In the limit where both the source and receiver are electrically large, i.e., $\frac{L_{\textrm{s}}}{\lambda} \rightarrow \infty$ and $\frac{L_{\textrm{r}}}{\lambda} \rightarrow \infty$, \eqref{eqn: Discrete NLoS} converges to the plane-wave model in \eqref{eqn: LoS Plane Wave} with the angular-domain channel impulse response in \eqref{eqn:AngularChannelNLoS}. This behavior is analogous to the Fourier series converging to the Fourier transform for a signal with angular bandwidth $\Omega$ and period $T$, as $\Omega T\rightarrow \infty$. Next, we obtain the channel matrix corresponding to $h^{\nlos}(\r,\s)$ by spatially sampling the linear regions at the source and receiver.

\smallskip

\textbf{\textit{\ac{NLoS} channel matrix.}} To represent \( h^{\nlos}(\r, \s) \) in terms of \ac{MIMO} channel, the Fourier series expansion in \eqref{eqn: Discrete NLoS} is spatially sampled, resulting in the approximate matrix representation~\cite{pizzo2022fourier}
\begin{align}
\label{eqn: ChannelMatrixSamplingNLoS}
\H^{\nlos}\approx\sqrt{N_{\textrm{r}} N_{\textrm{s}}}\sum_{q_x \in\setE_{\textrm{r}}}\sum_{p_x\in\setE_{\textrm{s}}} H_{\textrm{a}}(q_x, p_x)\a_{\textrm{r}}(q_x)\a_{\textrm{s}}^\herm(p_x),
\end{align}
where $\a_{\textrm{s}}(p_x) \in \mathbb{C}^{N_{\textrm{s}}}$ is the transmit plane-wave vector with entries $\frac{1}{\sqrt{N_{\textrm{s}}}}a^*_{\textrm{s}}(p_x,\s_v)$, for $v=1,\ldots,N_{\textrm{s}}$, and $\a_{\textrm{r}}(q_x) \in \mathbb{C}^{N_{\textrm{r}}}$ is the receive plane-wave vector with entries $\frac{1}{\sqrt{N_{\textrm{r}}}}a_{\textrm{r}}(q_x,\r_u)$, for $u=1,\ldots,N_{\textrm{r}}$. Let $\A_{\textrm{s}} \in \mathbb{C}^{N_{\textrm{s}}\times  n_{\textrm{s}}}$ and $\A_{\textrm{r}}\in \mathbb{C}^{N_{\textrm{r}}\times  n_{\textrm{r}}}$ be deterministic matrices constructed by selecting columns from a \ac{2D} inverse discrete Fourier transform matrix. These matrices are semi-unitary, satisfying $\A_{\textrm{s}}^\herm \A_{\textrm{s}} = \I_{n_{\textrm{s}}}$ and $\A_{\textrm{r}}^\herm \A_{\textrm{r}} = \I_{n_{\textrm{r}}}$. Additionally, we introduce the vectors $\gammab_{\textrm{s}}\in\Real^{n_\textrm{s}}$ and $\gammab_{\textrm{r}}\in\Real^{n_\textrm{r}}$ collecting the coefficients $\{ \gamma(p_x): p_x \in \setE_{\textrm{s}} \}$ and $\{ \gamma(q_x): q_x \in \setE_{\textrm{r}} \}$, respectively. Based on these definitions, \eqref{eqn: ChannelMatrixSamplingNLoS} can be expressed in a more compact form as 
\begin{align}
    \label{eqn:ChannelMatrixStochastic}
    \H^{\nlos}\approx \A_{\textrm{r}} \Tilde{\H} \A_{\textrm{s}}^\herm,
\end{align}
with $\Tilde{\H} = e^{j\diag(\gammab_{\textrm{r}})r_z} \H_{\textrm{a}} e^{-j\diag(\gammab_{\textrm{s}}) s_z} \in \Compl^{n_{\textrm{r}} \times n_{\textrm{s}}}$. Furthermore,
\begin{align}
    \label{eqn:AngularChannelMatrixStochastic}
    \H_{\textrm{a}} = \Sigmab\odot \W \in \Compl^{n_{\textrm{r}} \times n_{\textrm{s}}}
\end{align}
denotes the angular-domain channel matrix, where $\Sigmab \in \Real_+^{n_{\textrm{r}} \times n_{\textrm{s}}}$ contains the scaled standard deviations $\left\{\sqrt{N_{\textrm{r}} N_{\textrm{s}}} \sigma(q_x, p_x) : p_x \in \setE_{\textrm{s}}, q_x \in \setE_{\textrm{r}} \right\}$ and $\W \in \mathbb{C}^{n_{\textrm{r}} \times n_{\textrm{s}}}$ is a random matrix with \ac{i.i.d.} $\setN_{\mathbb{C}}(0,1)$ entries.
We demonstrate in Section~\ref{sec: Numerical Results} that the approximation in \eqref{eqn:ChannelMatrixStochastic} achieves excellent consistency with the theoretical model.

\smallskip

Having analyzed the \ac{LoS} and \ac{NLoS} channels separately in Section~\ref{sec: Line of Sight Channel Model} and in this section, respectively, we now characterize the general case with \ac{LoS}+\ac{NLoS} propagation.

\subsection{LoS+NLoS Channel Model}
\label{sec: LoS+NLoS Channel Model}

In this section, we present the \ac{EM}-based \ac{LoS}+\ac{NLoS} channel model for holographic lines. A stationary complex Gaussian random process $c(t)$ with zero mean and infinite time interval was considered in \cite{pizzo2022spatial} to characterize the \ac{NLoS} component. Here, we extend this approach by introducing a complex Gaussian random process $\Tilde{c}(t)$ with non-zero mean, thereby capturing the contribution from both the \ac{LoS} and \ac{NLoS} components.

Let $Z(\omega)$ and $\tilde{\mu} = \mathbb{E}\big[\Tilde{c}(t)\big]$ denote the integrated Fourier transform and mean, respectively, of $\Tilde{c}(t)$. To express the Fourier spectral representation of $\tilde{c}(t)$ in a form consistent with \cite[Eq.~63]{pizzo2022spatial}, we begin by assuming
\begin{align}
    \label{eqn: modiefied dZ(omwga)}
    \diff Z(\omega)=\big(\sqrt{2\pi}\tilde{\mu}\delta(\omega)+\sqrt{S(\omega)}W(\omega)\big)\diff \omega,
\end{align}
where $S(\omega)$ is the \ac{PSD} of $c(t)$ and $W(\omega)$ is a complex white Gaussian noise random process defined in the frequency domain. Using \eqref{eqn: modiefied dZ(omwga)} and \cite[Eq.~61]{pizzo2022spatial}, the Fourier spectral representation of $\Tilde{c}(t)$ is written as
\begin{align}
    \label{eqn: Fourier Spectral Representation_ht Random process with mean}
    \hspace{-1mm}\Tilde{c}(t)=\frac{1}{\sqrt{2\pi}}\int_{-\infty}^{\infty}\big(\sqrt{2\pi}\tilde{\mu}\delta(\omega)+\sqrt{S(\omega)}W(\omega)\big)e^{j\omega t} \diff \omega.
\end{align}
Moreover, the \ac{ACF} of $\Tilde{c}(t)$ is given by
\begin{align}
\begin{split}
\hspace{-1mm} \mathbb{E}\big[\Tilde{c}(t)\Tilde{c}^{*}(s)\big]= \, & \int_{\Real^2}|\tilde{\mu}|^2\delta(\omega)\delta(\lambda)e^{j(\omega t-\lambda s)}\diff \omega \diff \lambda\\&
    + \! \frac{1}{2\pi}\int_{\Real^2}S(\omega)\delta(\omega \! - \! \lambda)e^{j(\omega t-\lambda s)}\diff \omega \diff \lambda \\
    = \, & |\tilde{\mu}|^2+\frac{1}{2\pi}\int_{-\infty}^{\infty}S(\omega)e^{j\omega(t-s)}\diff \omega \\
    = \, & |\tilde{\mu}|^2    +\mathbb{E}\big[c(t)c^{*}(s)\big],
\end{split}
\label{eqn: Autocorrelation Function Derivation with mean2}
\end{align}
where $\mathbb{E}\big[c(t)c^{*}(s)\big]=\frac{1}{2\pi}\int_{-\infty}^{\infty}S(\omega)e^{j\omega(t-s)}\diff \omega$ is the \ac{ACF} of $c(t)$. Thus, \eqref{eqn: Autocorrelation Function Derivation with mean2} verifies the assumption in \eqref{eqn: modiefied dZ(omwga)}, as the \ac{ACF} of a random process with non-zero mean is indeed the sum of $|\tilde{\mu}|^2$ and the \ac{ACF} of the corresponding zero-mean process.

\begin{figure*}
\setcounter{equation}{55}
    \begin{align}
    \label{eqn: Channel H_nm with FT1}
     H^{\los}_{n, m}={\frac{\kappa\eta \sqrt{L_{\textrm{s}} L_{\textrm{r}}}}{4\pi}}\int_{-\kappa}^{\kappa}\mathrm{sinc}\bigg(\bigg(\frac{\kappa_x}{2\pi}-\frac{n-\frac{N-1}{2}}{L_{\textrm{r}}}\bigg)L_{\textrm{r}}\bigg) \frac{e^{j\sqrt{\kappa^2-\kappa_x^2}d}}{\sqrt{\kappa^2-\kappa_x^2}} \mathrm{sinc}\bigg(\bigg(\frac{\kappa_x}{2\pi}-\frac{m-\frac{N-1}{2}}{L_{\textrm{s}}}\bigg)L_{\textrm{s}}\bigg) \diff \kappa_x
\end{align}
\setcounter{equation}{44}
\hrule \vspace{-4mm}
\end{figure*}

Equivalent to the time-frequency domain representation in \eqref{eqn: Fourier Spectral Representation_ht Random process with mean}, the Fourier spectral representation of a spatially stationary complex Gaussian random channel $h(\r, \s)$ with non-zero mean is given in \eqref{eqn:LoS NLoS Spatial Random Process} at the top of the page, with $a_{\textrm{s}}(\kappab,\s)$ and $a_{\textrm{r}}(\k,\r)$ defined in \eqref{eqn:plane waves source and receiver}. Comparing \eqref{eqn:LoS NLoS Spatial Random Process} with \eqref{eqn: LoS Plane Wave} and \eqref{eqn:NLoSSpatial}, and using \eqref{eqn:PSD} together with $\tilde{\mu}=\frac{\kappa\eta}{2\gamma(\kappa_{x})}$, we write the \ac{LoS}+\ac{NLoS} channel impulse response as
\setcounter{equation}{39}
\begin{align}
    \label{eqn: General Channel Model}
    h(\r, \s)=h^{\los}(\r, \s)+h^{\nlos}(\r, \s).
\end{align}
The corresponding \ac{LoS}+\ac{NLoS} kernel, analogous to the \ac{NLoS} kernel in \eqref{eqn:NLoS_Kernel}, is given by
\begin{align}
    \label{eqn: Los+NLoS Kernel}
    \tilde{K}(k_x, \kappa_x)=\delta(k_x-\kappa_x)+A(k_x, \kappa_x)W(k_x, \kappa_x).
\end{align}
Moreover, the corresponding \ac{LoS}+\ac{NLoS} \ac{PSD}, obtained by applying the Fourier transform to \eqref{eqn: Autocorrelation Function Derivation with mean2} and expressing the result in the wavenumber domain, is given by
\begin{align}
    \label{eqn: PSD Los+NLoS}
    \tilde{S}(k_x,\kappa_x)=(2\pi)^2 |\tilde{\mu}|^2 \delta(k_x)\delta(\kappa_x) + S(k_x,\kappa_x).
\end{align}
Using the \ac{LoS} and \ac{NLoS} channel matrices derived in \eqref{eqn:Closed LoS Matrix} and \eqref{eqn:ChannelMatrixStochastic}, respectively, it is evident from \eqref{eqn: General Channel Model} that the \ac{LoS} and \ac{NLoS} components combine additively even in the near field. Therefore, the \ac{LoS}+\ac{NLoS} channel matrix is expressed as
\begin{align}
    \label{eqn: General Channel Matrix}
   \H=\H^{\los}+\H^{\nlos},
\end{align}
with spatial covariance matrix given by
\begin{align}
 \label{eqn:SpatialCorrelation}
    \hspace{-1mm} \R \! = \! \mathbb{E}\big[\vec(\H \! - \! \H^{\los})\vec(\H \! - \! \H^{\los})^\herm\big] \in\mathbb{C}^{N_{\textrm{r}}N_{\textrm{s}}\times N_{\textrm{r}}N_{\textrm{s}}}.
\end{align}

In the next section, we apply the \ac{WDM} framework to the \ac{EM}-based \ac{LoS}+\ac{NLoS} channel model for holographic lines developed so far.

\begin{figure}[t]
    \centering
    \begin{tikzpicture}[>=latex]

\small

\def\dh{0.75cm}
\def\dv{0.75cm}
\node[] (center) at (0,0) {};
\node[left=\dh of center] (center_left) {};
\node[above=\dv of center_left] (top_left) {};
\node[below=\dv of center_left] (bottom_left) {};
\node[right=\dh of center] (center_right) {};
\node[above=\dv of center_right] (top_right) {};
\node[below=\dv of center_right] (bottom_right) {};

\draw[very thick] (top_left) -- (bottom_left) node[midway, right=1pt] {$\j(\s)$};
\draw[very thick] (top_right) -- (bottom_right) node[midway, left=1pt] {$\e(\r)$};

\node[draw, minimum width=0.8cm, minimum height=0.5cm, fill=gray!20, ellipse, rotate=60, transform shape] () at ($(center)+(0,\dv)$) {};
\node[draw, minimum width=0.5cm, minimum height=0.4cm, fill=gray!20, ellipse, rotate=120, transform shape] () at ($(center)+(0,-\dv)$) {};

\node[draw, circle, left=\dh/4 of center_left, align=center, inner sep=0pt] (+_left) {+};
\draw[->] (+_left.east) -- (center_left.center);
\node[right=\dh/4 of center_right, align=center, inner sep=0pt] (cross_right) {};
\draw[-] (center_right.center) -- (cross_right.center);

\node[draw, rectangle, above left=\dv and \dh/3 of +_left.center, align=center, text width=1cm] (phi_1) {$\phi_1(s_x)$};
\node[draw, rectangle, below left=\dv and \dh/3 of +_left.center, align=center, text width=1cm] (phi_2) {$\phi_N(s_x)$};
\node[align=center] () at ($(+_left)+(-\dv-0.1cm,0cm)$) {$\vdots$};
\draw[->] (phi_1.east) -- ($(phi_1.east)+(\dh/3,0)$) -- (+_left.north);
\draw[->] (phi_2.east) -- ($(phi_2.east)+(\dh/3,0)$) -- (+_left.south);

\node[draw, rectangle, above right=\dv and \dh/3 of cross_right.center, align=center, text width=2cm] (psi_1) {$\int(\cdot)\psi_{1}^{*}(r_x)\diff r_x$};
\node[draw, rectangle, below right=\dv and \dh/3 of cross_right.center, align=center, text width=2cm] (psi_2) {$\int (\cdot)\psi_{N}^{*}(r_x)\diff r_x$};
\node[align=center] () at ($(cross_right)+(\dv+0.6cm,0cm)$) {$\vdots$};
\draw[->] (cross_right.center) -- ($(psi_1.west)+(-\dh/3,0)$) -- (psi_1.west);
\draw[->] (cross_right.center) -- ($(psi_2.west)+(-\dh/3,0)$) -- (psi_2.west);

\node[left=\dv/2 of phi_1, align=center, inner sep=0pt] (x_1) {$\bar{x}_1$};
\node[left=\dv/2 of phi_2, align=center, inner sep=0pt] (x_2) {$\bar{x}_N$};
\draw[->] (x_1.east) -- (phi_1.west);
\draw[->] (x_2.east) -- (phi_2.west);

\node[draw, circle, right=\dv/2 of psi_1, align=center, inner sep=0pt] (+_right_1) {+};
\node[draw, circle, right=\dv/2 of psi_2, align=center, inner sep=0pt] (+_right_2) {+};
\node[right=\dv/2 of +_right_1, align=center, inner sep=0pt] (y_1) {$y_1$};
\node[right=\dv/2 of +_right_2, align=center, inner sep=0pt] (y_2) {$y_N$};
\node[below=\dv/2 of +_right_1, align=center, inner sep=0pt] (z_1) {$z_1$};
\node[above=\dv/2 of +_right_2, align=center, inner sep=0pt] (z_2) {$z_N$};
\draw[->] (psi_1.east) -- (+_right_1.west);
\draw[->] (psi_2.east) -- (+_right_2.west);
\draw[->] (+_right_1.east) -- (y_1.west);
\draw[->] (+_right_2.east) -- (y_2.west);
\draw[->] (z_1.north) -- (+_right_1.south);
\draw[->] (z_2.south) -- (+_right_2.north);

\draw[decorate, decoration={brace, amplitude=5pt}] ($(bottom_right)+(0,0mm)$) -- ($(bottom_left)+(0,0mm)$) node[midway, yshift=-3.5mm] {\footnotesize EM-based channel};
\draw[decorate, decoration={brace, amplitude=5pt}] ($(bottom_right)+(27.6mm,-6mm)$) -- ($(bottom_left)+(-18.5mm,-6mm)$) node[midway, yshift=-3.5mm] {\footnotesize WDM-applied channel};

\end{tikzpicture}
    \caption{A schematic of the considered \ac{WDM}-applied \ac{LoS}+\ac{NLoS} holographic \ac{MIMO} system model.}
    \label{fig:WDM System Model}
\end{figure}

\section{WDM with LoS+NLoS Channel}
\label{sec: Wavenumber-Division Multiplexing}

The \ac{WDM} framework was originally proposed in \cite{sanguinetti2022wavenumber} for the purely \ac{LoS} scenario. Building on its core concepts, we extend it here to the \ac{LoS}+\ac{NLoS} channel model for holographic lines developed in Section~\ref{sec:holographicMIMO}, and analyze the resulting channel formulation. As detailed in Section~\ref{sec: LoS+NLoS Channel Model}, the \ac{LoS}+\ac{NLoS} channel is the sum of its \ac{LoS} and \ac{NLoS} components. Accordingly, after outlining the general \ac{WDM} framework, we study the \ac{WDM}-applied \ac{LoS} and \ac{NLoS} scenarios separately in Sections~\ref{sec: WDM for LoS} and~\ref{sec: WDM for NLoS}, respectively, and obtain the \ac{WDM}-applied \ac{LoS}+\ac{NLoS} channel by summing the two. Lastly, we derive the spatial autocorrelation properties of the \ac{WDM}-applied \ac{NLoS} channel in Section~\ref{sec: Autocorrelation of Channel H_{nm}}.

Consider a \ac{WDM}-applied holographic \ac{MIMO} system with \ac{LoS}+\ac{NLoS} channel as depicted in Fig.~\ref{fig:WDM System Model}. For simplicity we set $s_z=0$ and $r_z=d$; consequently, the source and receiver span the linear regions $\setL_{\textrm{s}}=\{(s_x, 0): |s_x|\leq \frac{L_{\textrm{s}}}{2}\} $ and $\setL_{\textrm{r}}= \{(r_x, d): |r_x|\leq \frac{L_{\textrm{r}}}{2}\}$, respectively. Let $\{\bar{x}_m\}_{m=1}^{N}$ denote the transmitted data symbols, with $N \leq \mathrm{min}(n_{\textrm{s}}, n_{\textrm{r}})$. The electric current at the source (measured in amperes) is constructed as $i(s_x) = \sum_{m=1}^N \bar{x}_m \phi_m(s_x)$, where $\big\{\phi_m(s_x)\big\}_{m=1}^{N}$ represent the transmit Fourier basis, with $m$-th basis function
\begin{align}
\label{eqn: basis function at source}
   \phi_m(s_x)= 
\begin{cases}
    \frac{1}{\sqrt{L_{\textrm{s}}}}e^{j\frac{2\pi}{L_{\textrm{s}}}m s_x},&  |s_x|\leq \frac{L_{\textrm{s}}}{2}\\
    0,              & \textrm{otherwise},
\end{cases}
\end{align}
and with $\bar{x}_m=\int_{-\frac{L_{\textrm{s}}}{2}}^{\frac{L_{\textrm{s}}}{2}} i(s_x)\phi_m^*(s_x) ds_x$. The current density at the source is defined as
\begin{align}
\label{eqn: current distribution}
   \j(\s)=i(s_x)\delta(s_z)\hat{\x},
\end{align}
where $\hat{\x} = [1, 0, 0]^\tran$ is the unit vector along the $x$-axis.
The electric field at the receive point $\r$ given the current density at the source in \eqref{eqn: current distribution} is expressed as \cite{pizzo2020holographic}
\begin{align}
    \label{eqn: convolution spatially variant}
    \e(\r)=\int_{-\infty}^{\infty} h(\r, \s)\,\j(\s)\diff\s,
\end{align}
with $h(\r, \s)$ defined in \eqref{eqn: General Channel Model}. Using \eqref{eqn: current distribution} and \eqref{eqn: convolution spatially variant}, we have 
\begin{align}
    \label{eqn: convolution spatially variant with current distribution}
\hspace{-1.4mm}\e(\r)\!=\!\int_{-\frac{L_{\textrm{s}}}{2}}^{\frac{L_{\textrm{s}}}{2}}\! h(\r, \s)\! \sum_{m=1}^N \!\xi_m\phi_m(s_x) \hat{\x} \diff s_x\! = \!\big[e_{x}(r_{x}), 0, 0 \big]^{\tran}.\!
\end{align}
Then, $\e(\r)$ is projected onto the inner-product space spanned~by
\begin{align}
    \label{eqn: receiver vector space}
    \psib_n(\r)= \psi_n(r_x)\delta(r_z-d)\hat{\x},
\end{align}
where $\big\{\psi_n(r_x)\big\}_{n=1}^{N}$ represent the receive Fourier basis, with $n$-th basis function
\begin{align}
\label{eqn: basis function at receiver}
\psi_n(r_x)=
\begin{cases}
    \frac{1}{\sqrt{L_{\textrm{r}}}}e^{j\frac{2\pi}{L_{\textrm{r}}}n r_x},&  |r_x|\leq \frac{L_{\textrm{r}}}{2}\\
    0,              & \textrm{otherwise}.
\end{cases}
\end{align}
Hence, the $n$-th spatial sample of the received signal is given~by
\begin{align}
\label{eqn: spatial sample at receiver}
  y_n=\int_{-\frac{L_{\textrm{r}}}{2}}^{\frac{L_{\textrm{r}}}{2}} \psi_n^*(r_x) e_{x}(r_{x}) \diff r_x+ z_n,
\end{align}
where $z_n \in \setN_{\Compl}(0, \chi^2)$ is the \ac{AWGN} term with variance $\chi^2$. Finally, plugging \eqref{eqn: convolution spatially variant with current distribution} into \eqref{eqn: spatial sample at receiver} yields
 \begin{align}
\label{eqn: input-output relationship}
  y_n=\sum_{m=1}^{N} H_{n, m} \bar{x}_m+ z_n,\quad n=1,...,N,
\end{align}
where
\begin{align}
\label{eqn: Channel H_nm mod}
 H_{n, m}=\int_{-\frac{L_{\textrm{r}}}{2}}^{\frac{L_{\textrm{r}}}{2}}\int_{-\frac{L_{\textrm{s}}}{2}}^{\frac{L_{\textrm{s}}}{2}}\psi_n^*(r_x) h(\r, \s) \phi_m(s_x) \diff s_x\diff r_x
\end{align}
is the \ac{WDM}-applied \ac{LoS}+\ac{NLoS} channel corresponding to the $m$-th transmit and $n$-th receive Fourier basis functions.\footnote{Although we employ Fourier bases as in the original \ac{WDM} framework, other orthogonal bases may also be used \cite{miller2000communicating}.} Having established \eqref{eqn: Channel H_nm mod}, we now analyze the \ac{WDM}-applied \ac{LoS}~scenario.

\subsection{WDM with LoS Channel}
\label{sec: WDM for LoS}

In this section, we present the \ac{WDM}-applied \ac{LoS} channel, denoted by $H^{\los}_{n,m}$. A corresponding expression was previously derived in \cite[Eq.~(62)]{sanguinetti2022wavenumber} using the vector Green’s function. In contrast, we adopt the scalar Green’s function, enabling a direct comparison with the ray-tracing-based model in \eqref{eqn:Ray Tracing Channel Gain} and the \ac{EM}-based \ac{LoS} channel model in \eqref{eqn:Closed LoS Matrix} for holographic lines; nonetheless, as demonstrated in Section~\ref{sec: Numerical Results}, the formulations with vector and scalar Green's functions yield nearly identical results. Applying the general \ac{WDM} framework in \eqref{eqn: Channel H_nm mod} to \eqref{eqn:Closed LoS} yields
\begin{align}
    \label{eqn: WDM LoS H_nm}
    H^{\los}_{n, m}=\,& \frac{\kappa\eta}{4}\int_{-\frac{L_{\textrm{r}}}{2}}^{\frac{L_{\textrm{r}}}{2}}\int_{-\frac{L_{\textrm{s}}}{2}}^{\frac{L_{\textrm{s}}}{2}}\psi_n^*(r_x) H_{0}^{(1)}\big(\kappa\|\r-\s\|\big)\nonumber \\ & \times \phi_m(s_x) \diff s_x\diff r_x.
\end{align}
An equivalent representation of \eqref{eqn: WDM LoS H_nm} can be obtained based on \eqref{eqn: Hankel expansion} as
\begin{align}
    \label{eqn: Channel H_nm with FT}
    H^{\los}_{n, m}={\frac{\kappa\eta}{4\pi}}\int_{-\infty}^{\infty}\Psi_n^*(\kappa_x) \frac{e^{j\gamma(\kappa_x)d}}{\gamma(\kappa_x)} \Phi_m(\kappa_x) \diff \kappa_x,
\end{align}
where $\Psi_n(\kappa_x)$ and $\Phi_m(\kappa_x)$ denote the Fourier transforms of $\psi_n(r_x)$ and $\phi_m(s_x)$, respectively. Finally, explicitly expressing $\Psi_n(\kappa_x)$ and $\Phi_m(\kappa_x)$ leads to \eqref{eqn: Channel H_nm with FT1} at the top of the page. Since deriving \eqref{eqn: Channel H_nm with FT1} in closed form is challenging, we evaluate it numerically in Section~\ref{sec: Numerical Results}.

\subsection{WDM with NLoS Channel}
\label{sec: WDM for NLoS}

In this section, we present the \ac{WDM}-applied \ac{NLoS} channel, denoted by $H^{\nlos}_{n,m}$. Combining \eqref{eqn: Channel H_nm mod} and \eqref{eqn:NLoSSpatial} leads to
\setcounter{equation}{56}
\begin{align}
\label{eqn: Channel H_nm}
 H^{\nlos}_{n, m}=\, &  \frac{1}{2\pi}\int_{-\frac{L_{\textrm{r}}}{2}}^{\frac{L_{\textrm{r}}}{2}}\int_{-\frac{L_{\textrm{s}}}{2}}^{\frac{L_{\textrm{s}}}{2}}\int_{\Real^2}\psi_n^*(r_x)a_{\textrm{r}}(\k,\r) \frac{A(k_x, \kappa_x)}{\sqrt{\gamma(k_x)}}\nonumber  \\ &\times \frac{W(k_x, \kappa_x)}{\sqrt{\gamma(\kappa_x)}} a_{\textrm{s}}(\kappab,\s) \phi_m(s_x) \diff k_x\diff\kappa_x \diff s_x\diff r_x,
\end{align}
which is used in Section~\ref{sec: Autocorrelation of Channel H_{nm}} to derive its spatial autocorrelation. Now, we substitute the Fourier series expansion of $h^{\nlos}(\r, \s)$, given in \eqref{eqn: Discrete NLoS} with angular-domain channel impulse response in \eqref{eqn: Discrete Angular}, into \eqref{eqn: Channel H_nm mod}. Utilizing \eqref{eqn: Discretize Plane wave source and receiver} along with the fact that $H_{\textrm{a}}(q_x, p_x)$ and $e^{j\frac{2\pi}{L_{\textrm{r}}}\gamma(q_x)r_z} H_{\textrm{a}}(q_x, p_x)e^{-j\frac{2\pi}{L_{\textrm{s}}}\gamma(p_x)s_z}$ are statistically equivalent, $H^{\nlos}_{n, m}$ can be expressed with a slight abuse of notation as
\begin{align}
\begin{split}
 H^{\nlos}_{n, m}=\, & \frac{1}{\sqrt{L_{\textrm{r}} L_{\textrm{s}}}}\int_{-\frac{L_{\textrm{r}}}{2}}^{\frac{L_{\textrm{r}}}{2}}\int_{-\frac{L_{\textrm{s}}}{2}}^{\frac{L_{\textrm{s}}}{2}}\sum_{q_x\in\setE_{\textrm{r}}}\sum_{p_x\in\setE_{\textrm{s}}} H_{\textrm{a}}({q_x, p_x})\\
 &\times e^{j\frac{2\pi}{L_{\textrm{r}}}(q_x-n) r_x}e^{-j\frac{2\pi}{L_{\textrm{s}}} (p_x-m) s_x} \diff s_x\diff r_x \\
 = \, & \sqrt{L_{\textrm{r}} L_{\textrm{s}}}\sum_{q_x\in\setE_{\textrm{r}}}\sum_{p_x \in\setE_{\textrm{s}}}H_{\textrm{a}}({q_x, p_x})\mathrm{sinc}(q_x-n) \\
 &\times \mathrm{sinc}(p_x-m) \\
 = \, & \sqrt{L_{\textrm{r}} L_{\textrm{s}}}H_{\textrm{a}}({n, m}), 
\end{split}
\label{eqn: Channel H_nm discrete without sinc}
\end{align}
with $n\in\setE_{\textrm{r}}$ and $m\in\setE_{\textrm{s}}$. From \eqref{eqn: Channel H_nm discrete without sinc}, we observe that the channel resulting from applying \ac{WDM} corresponds to the angular-domain channel scaled by a gain that depends on the lengths of the holographic lines. This result holds primarily under two key conditions: the existence of a Fourier plane-wave series expansion of the \ac{NLoS} channel, enabled by modeling $h^{\nlos}(\r, \s)$ as a spatially stationary Gaussian random field, and the use of Fourier basis functions in the \ac{WDM} framework. On the one hand, the Fourier plane-wave series expansion identifies the wavenumber harmonics present in the scattered field. On the other hand, the use of Fourier bases in the \ac{WDM} framework can be interpreted as applying a Fourier transform in the spatial-frequency domain, thereby revealing the underlying wavenumber spectrum and, consequently, the related angular-domain channel (see footnote~\ref{fn:ang_wave}). Having established the \ac{WDM}-applied \ac{NLoS} channel, we next analyze its spatial autocorrelation properties.

\subsection{Spatial Autocorrelation of \texorpdfstring{$H^{\nlos}_{n, m}$}{}}
\label{sec: Autocorrelation of Channel H_{nm}}

In this section, we first derive the spatial autocorrelation of the \ac{WDM}-applied \ac{NLoS} channel and then establish the relation between its \ac{PSD} and \ac{PSF}. This relation is analogous to the one in \eqref{eqn:PSD}, which was obtained by expressing \eqref{eqn:NLoSSpatial} in the form of \eqref{eqn:NLoSSpatialRandomProcess} following the correspondence between the Fourier spectral representation of a random process and the plane-wave representation in \ac{EM} theory. The spatial autocorrelation of the \ac{WDM}-applied \ac{NLoS} channel is given by
\begin{align}
    \label{eqn: Autocorrelation}
    R_{mn, qp}=\mathbb{E}\big[ H^{\nlos}_{n, m} (H^{\nlos}_{p, q})^*\big].
\end{align}
Plugging \eqref{eqn: Channel H_nm} into \eqref{eqn: Autocorrelation} leads to
\begin{align}
\label{eqn: Autocorrelation1}
  R_{mn, qp}=\,&\frac{1}{(2\pi)^2} \mathbb{E} \biggl[\int_{-\frac{L_{\textrm{r}}}{2}}^{\frac{L_{\textrm{r}}}{2}}\int_{-\frac{L_{\textrm{s}}}{2}}^{\frac{L_{\textrm{s}}}{2}}\int_{\Real^2}\psi_n^*(r_x)a_{\textrm{r}}(\k,\r)
 \frac{A(k_x, \kappa_x)}{\sqrt{\gamma(k_x)}}\nonumber  \\ &\times \frac{W(k_x, \kappa_x)}{\sqrt{\gamma(\kappa_x)}} a_{\textrm{s}}(\kappab,\s) \phi_m(s_x) \diff k_x\diff\kappa_x  \diff s_x\diff r_x \nonumber \\ 
 &\times \int_{-\frac{L_{\textrm{r}}}{2}}^{\frac{L_{\textrm{r}}}{2}}\int_{-\frac{L_{\textrm{s}}}{2}}^{\frac{L_{\textrm{s}}}{2}}\int_{\Real^2}\psi_p(r'_x)a^*_{\textrm{r}}(\k',\r') \frac{A(k'_x, \kappa'_x)}{\sqrt{\gamma(k'_x)}} \nonumber \\
 &\times \frac{W^*(k'_x, \kappa'_x)}{\sqrt{\gamma(\kappa'_x)}}a^*_{\textrm{s}}(\kappab',\s')\phi^*_q(s'_x) \diff k'_x\diff \kappa'_x  \diff s'_x\diff r'_x\biggr].
\end{align}
Since $W(k_x, \kappa_x)$ is a spatially stationary complex white Gaussian noise random field, we have
\begin{align}
    \label{eqn: Autocorrelation White noise}
    \mathbb{E}\big[W(k_x, \kappa_x)W^*(k'_x, \kappa'_x)\big]=\delta(k_x-k'_x)\delta(\kappa_x-\kappa'_x).
\end{align}
Moreover, $W(k_x, \kappa_x)$ and $e^{j\frac{2\pi}{L_{\textrm{r}}}\gamma(k_x)r_z} W(k_x, \kappa_x)e^{-j\frac{2\pi}{L_{\textrm{s}}}\gamma(\kappa_x)s_z}$ are statistically equivalent, which allows to express \eqref{eqn: Autocorrelation1} as
\begin{align}
\label{eqn: Autocorrelation2}
 R_{mn, qp}=\,& \frac{1}{(2\pi)^2} \int_{-\frac{L_{\textrm{r}}}{2}}^{\frac{L_{\textrm{r}}}{2}}\int_{-\frac{L_{\textrm{s}}}{2}}^{\frac{L_{\textrm{s}}}{2}}\int_{-\frac{L_{\textrm{r}}}{2}}^{\frac{L_{\textrm{r}}}{2}}\int_{-\frac{L_{\textrm{s}}}{2}}^{\frac{L_{\textrm{s}}}{2}}\int_{\Real^2} \psi_n^*(r_x)\nonumber\\ &\times e^{jk_x(r_x-r'_x)}
 \psi_p(r'_x) \frac{A^2(k_x, \kappa_x)}{\gamma(k_x)\gamma(\kappa_x)}e^{-j\kappa_x (s_x-s'_x)}\nonumber \\ &\times \phi_m(s_x) 
  \phi^*_q(s'_x)\diff k_x\diff\kappa_x   \diff s_x\diff r_x \diff s'_x\diff r'_x.
\end{align}

\begin{remark} \label{rem:separable_scattering}
To the best of our knowledge, an analytical functional characterization of $A^2(k_x, \kappa_x)$ that captures non-separable scattering effects at both the source and receiver is not available in the existing literature. Therefore, from this point onward, we assume separable scattering, which yields a Kronecker-structured \ac{NLoS} channel. Nevertheless, assuming that the joint \ac{PSF} $A^2(k_x, \kappa_x)$ were available, the subsequent analysis could be extended to non-separable scattering, albeit at the cost of increased analytical complexity in deriving the closed-form expressions of the \ac{ACF} and \ac{PSD} in Section~\ref{sec: Numerical Analysis}.
\end{remark}

Considering separable scattering at the source and receiver \cite{pizzo2022fourier}, such that $A^2(k_x, \kappa_x) \! = \! A_\textrm{r}^2(k_x)A_{\textrm{s}}^2(\kappa_x)$, we express \eqref{eqn: Autocorrelation2}~as
\begin{align}
\label{eqn: Autocorrelation Matrix elements}
[\bar{\R}]_{mn, qp}=[\bar{\R}_{\textrm{s}}]_{m, q} [\bar{\R}_{\textrm{r}}]_{n, p},
\end{align}
where
\begin{subequations}
\begin{align}
\label{eqn: Autocorrelation Matrix Source}
\big[\bar{\R}_{\textrm{s}}\big]_{m, q} =\,& \frac{1}{2\pi}\int_{-\frac{L_{\textrm{s}}}{2}}^{\frac{L_{\textrm{s}}}{2}}\int_{-\frac{L_{\textrm{s}}}{2}}^{\frac{L_{\textrm{s}}}{2}}\int_{-\infty}^{\infty}\phi_m(s_x)\frac{A_{\textrm{s}}^2( \kappa_x)}{\gamma(\kappa_x)}e^{-j\kappa_x (s_x-s'_x)}\nonumber\\
&\times \phi^*_q(s'_x)\diff\kappa_x\diff s_x\diff s'_x, \\
\label{eqn: Autocorrelation Matrix Receiver}
 \big[\bar{\R}_{\textrm{r}}\big]_{n, p}=\,& \frac{1}{2\pi}\int_{-\frac{L_{\textrm{r}}}{2}}^{\frac{L_{\textrm{r}}}{2}}\int_{-\frac{L_{\textrm{r}}}{2}}^{\frac{L_{\textrm{r}}}{2}}\int_{-\infty}^{\infty}\psi_n^*(r_x)\frac{A_{\textrm{r}}^2(k_x)}{\gamma(k_x)}e^{jk_x(r_x-r'_x)}\nonumber\\
 &\times \psi_p(r'_x)  \diff k_x \diff r_x\diff r'_x
\end{align}
\end{subequations}
are the transmit and receive spatial autocorrelations, respectively. Hence, \eqref{eqn: Autocorrelation Matrix elements} can be written in matrix form as
\begin{align}
\label{eqn: Autocorrelation Matrix}
 \bar{\R}=\bar{\R}_{\textrm{s}}\otimes \bar{\R}_{\textrm{r}}\in \mathbb{C}^{n_\textrm{r} n_\textrm{s}\times n_\textrm{r} n_\textrm{s}},
\end{align}
where $\bar{\R}_{\mathrm{s}} \in \mathbb{C}^{n_{\mathrm{s}} \times n_{\mathrm{s}}}$ and $\bar{\R}_{\mathrm{r}} \in \mathbb{C}^{n_{\mathrm{r}} \times n_{\mathrm{r}}}$ denote the spatial autocorrelation matrices at the source and receiver, respectively. Comparing \eqref{eqn: Autocorrelation Matrix Receiver} with the expression in \cite[Eq.~(71)]{sanguinetti2022wavenumber}, i.e.,
\begin{align}
    \label{eqn: WDM ACM}
    \hspace{-2mm} \big[\bar{\R}_{\textrm{r}}\big]_{n, p}=\int_{-\frac{L_{\textrm{r}}}{2}}^{\frac{L_{\textrm{r}}}{2}}\int_{-\frac{L_{\textrm{r}}}{2}}^{\frac{L_{\textrm{r}}}{2}}\Gamma_{\textrm{r}}(r_x-r'_x)\psi_n^*(r_x)\psi_p(r'_x) \diff r_x\diff r'_x,
\end{align}
we have the \ac{ACF} at the receiver given by
\begin{align}
\begin{split}
    \Gamma_{\textrm{r}}(r_x) & =\frac{1}{2\pi}\int_{-\infty}^{\infty}\frac{A_{\textrm{r}}^2(k_x)}{\gamma(k_x)}e^{jk_x r_x} \diff k_x \\
     & =\frac{1}{2\pi}\int_{-k}^{k}\frac{A_{\textrm{r}}^2(k_x)}{\sqrt{k^2-k_x^2}}e^{jk_x r_x} \diff k_x,
\end{split}
\label{eqn: autocorrelation1}
\end{align}
where the last equality follows from \eqref{eqn:gamma} for $k_x\in \setD$.

\begin{remark}
Under the assumption of separable scattering, the \ac{ACF} in \eqref{eqn: autocorrelation1} can also be derived by substituting \eqref{eqn:NLoSSpatial} into
\begin{align}
    \label{eqn: acf using h(r,s)}
    \Gamma(\r, \s)=\mathbb{E}\big[h(\r', \s')h^{*}(\r'+\r, \s'+\s)\big],
\end{align}
and using \eqref{eqn: Autocorrelation White noise} to simplify the otherwise complicated four-dimensional integration to a \ac{2D} one. Then, utilizing the separability assumption, $\Gamma(\r, \s)$ decomposes into the corresponding \acp{ACF} at the source and receiver.
\end{remark}

\noindent Applying the change of variable $k_x \! = \! k\cos\theta_{\textrm{r}}$, \eqref{eqn: autocorrelation1} simplifies~to
\begin{align}
    \label{eqn: autocorrelation2}    
    \Gamma_{\textrm{r}}(r_x)=\frac{1}{2\pi}\int_{0}^{\pi}A_{\textrm{r}}^2(\theta_{\textrm{r}})e^{jk\cos\theta_{\textrm{r}} r_x} \diff \theta_{\textrm{r}},
\end{align}
which is utilized in Section~\ref{sec: Numerical Analysis} to derive a closed-form expression for the \ac{ACF} and \ac{PSD}. The \ac{ACF} and \ac{PSD} are related through the Fourier transform in the wavenumber domain as
\begin{align}
    \label{eqn: Fourier Transform}
    S_{\textrm{r}}(k_x)=\int_{-\infty}^{\infty}\Gamma_{\textrm{r}}(r_x)e^{-jk_x r_x} \diff r_x,
\end{align}
where $S_{\textrm{r}}(k_x)$ is the \ac{PSD} at the receiver. Plugging \eqref{eqn: autocorrelation1} into \eqref{eqn: Fourier Transform} and rearranging the order of integration yields
\begin{align}
    \label{eqn: Fourier Transform1}
    S_{\textrm{r}}(k_x)=\frac{1}{2\pi}\int_{-k}^{k}\frac{A_{\textrm{r}}^2(k'_x)}{\gamma(k'_x)}\diff k'_x\int_{-\infty}^{\infty}e^{j(k'_x-k_x)r_x} \diff r_x.
\end{align}
Now, using again the relation $\int_{-\infty}^{\infty}e^{j(k'_x-k_x)r_x} \diff r_x=2\pi\delta(k'_x-k_x)$, \eqref{eqn: Fourier Transform1} simplifies to
\begin{align}
    \label{eqn: Fourier Transform2}
    S_{\textrm{r}}(k_x)=\frac{A_{\textrm{r}}^2(k_x)}{\gamma(k_x)} \mathbbm{1}_{\setD}(k_{x}),
\end{align}
which connects the \ac{PSD} and \ac{PSF}. Note that \eqref{eqn: Fourier Transform2} is consistent with \eqref{eqn:PSD} under the assumption of separable scattering. Under this assumption, we have $\sigma^2(q_x, p_x)=\sigma_{\textrm{s}}^2(p_x)\sigma_{\textrm{r}}^2(q_x)$, where $\sigma_{\textrm{s}}^2(p_x)$ and $\sigma_{\textrm{r}}^2(q_x)$ are the transmit and receive variances, respectively. Therefore, employing \eqref{eqn: Channel H_nm discrete without sinc}--\eqref{eqn: Autocorrelation}, we express the spatial autocorrelation as
\begin{align}
\label{eqn: Autocorrelation angular1}
    R_{mn, qp}= L_{\textrm{s}} L_{\textrm{r}} \sigma_{\textrm{r}}^2(q_x)\sigma_{\textrm{s}}^2(p_x)\delta[q_x-q_x']\delta[p_x-p_x'].
\end{align}
Then, \eqref{eqn: Autocorrelation angular1} can be written in matrix form as
\begin{align}
\label{eqn: Autocorrelation Matrix different}
 \bar{\R}= \diag(\bar{\sigmab}_{\textrm{s}} \odot \bar{\sigmab}_{\textrm{s}})\otimes \diag(\bar{\sigmab}_{\textrm{r}} \odot \bar{\sigmab}_{\textrm{r}}),
\end{align}
where $\bar{\sigmab}_{\textrm{s}}\in \Real^{n_{\textrm{s}}}_+$ and $\bar{\sigmab}_{\textrm{r}}\in \Real^{n_{\textrm{r}}}_+$ collect the scaled standard deviations $\left\{\sqrt{L_{\textrm{s}}}\sigma_{\textrm{s}}(p_x): p_x\in \setE_{\textrm{s}}\right\}$ and $\left\{\sqrt{L_{\textrm{r}}}\sigma_{\textrm{r}}(q_x): q_x\in \setE_{\textrm{r}}\right\}$, respectively.
Now, comparing \eqref{eqn: Autocorrelation Matrix} with \eqref{eqn: Autocorrelation Matrix different}, we have
\begin{subequations}
\label{eqn: Autocorrelation Matrix Source Diff and Receiver Diff}
    \begin{align}
 \bar{\R}_{\textrm{s}}&=\diag(\bar{\sigmab}_{\textrm{s}} \odot \bar{\sigmab}_{\textrm{s}}),\\
 \bar{\R}_{\textrm{r}}&=\diag(\bar{\sigmab}_{\textrm{r}} \odot \bar{\sigmab}_{\textrm{r}}).
\end{align}
\end{subequations}

Lastly, adopting the structure of the transmit and receive spatial autocorrelation matrices presented in \cite{pizzo2022fourier}, we obtain the same matrices for the \ac{EM}-based \ac{NLoS} channel in \eqref{eqn:ChannelMatrixStochastic} as
\begin{subequations}
\label{eqn: Autocorrelation Matrix Source Diff and Receiver Diff Holographic}
\begin{align}
\R_{\textrm{s}}&=\A_{\textrm{s}}\diag(\sigmab_{\textrm{s}} \odot \sigmab_{\textrm{s}})\A^\herm_{\textrm{s}}\in \mathbb{C}^{N_{\mathrm{s}} \times N_{\mathrm{s}}},\\
\R_{\textrm{r}}&=\A_{\textrm{r}}\diag(\sigmab_{\textrm{r}} \odot \sigmab_{\textrm{r}})\A^\herm_{\textrm{r}}\in \mathbb{C}^{N_{\mathrm{r}} \times N_{\mathrm{r}}},
\end{align}
\end{subequations}
respectively, where $\sigmab_{\textrm{s}}\in\Real_+^{n_{\textrm{s}}}$ and $\sigmab_{\textrm{r}}\in\Real_+^{n_{\textrm{r}}}$ collect the scaled standard deviations $\left\{\sqrt{N_{\textrm{s}}}\sigma_{\textrm{s}}(p_x): p_x\in \setE_{\textrm{s}}\right\}$ and $\left\{\sqrt{N_{\textrm{r}}}\sigma_{\textrm{r}}(q_x): q_x\in \setE_{\textrm{r}}\right\}$, respectively.\footnote{Note that, from \eqref{eqn: Autocorrelation Matrix Source Diff and Receiver Diff Holographic}, we have $\Exp \big[ (\H^{\nlos})^{\herm} \H^{\nlos} \big] = \tr (\R_{\textrm{r}}) \R_{\textrm{s}}$ and $\Exp \big[ \H^{\nlos} (\H^{\nlos})^{\herm} \big] = \tr (\R_{\textrm{s}}) \R_{\textrm{r}}$.} Then, analogous to \eqref{eqn: Autocorrelation Matrix}, the spatial covariance matrix in \eqref{eqn:SpatialCorrelation} can be written as $\R=\R_{\textrm{s}}\otimes \R_{\textrm{r}}$ for separable scattering. Note that the eigenvalues of \eqref{eqn: Autocorrelation Matrix Source Diff and Receiver Diff} and \eqref{eqn: Autocorrelation Matrix Source Diff and Receiver Diff Holographic} differ solely by a multiplicative factor, i.e., their spectra are identical up to a constant scaling.

So far, we have developed the \ac{EM}-based channel model for holographic lines with \ac{LoS}+\ac{NLoS} propagation and analyzed it through the lens of \ac{WDM}. Next, we characterize the angular-domain \ac{NLoS} channel, which is required to complete the analysis of the \ac{EM}-based \ac{NLoS} channel in Section~\ref{sec: NLoS} and the \ac{WDM}-applied \ac{NLoS} channel in Section~\ref{sec: WDM for NLoS}.

\section{Angular-Domain NLoS Channel}
\label{sec: Calculation of Channel Coefficients for WDM and holographic channel}

In this section, we first characterize $H_{\textrm{a}}(q_x, p_x)$ and then obtain closed-form expressions for the \ac{ACF} and \ac{PSD} in the presence of isotropic and non-isotropic scattering.

\subsection{Characterization of \texorpdfstring{$H_{\mathrm{a}}(q_x, p_x)$}{}}
\label{sec: Modeling of H_a}

Characterizing $H_{\textrm{a}}(q_x, p_x)$ is essential for deriving expressions for the \ac{EM}-based and \ac{WDM}-applied \ac{NLoS} channel matrices. Since we have $H_{\textrm{a}}(q_x, p_x) \sim \setN_{\mathbb{C}}(0,\sigma^2(q_x, p_x))$ from \eqref{eqn: Discrete Angular}, the following analysis focuses on characterizing $\sigma^2(q_x, p_x)$.
Equation \eqref{eqn: Discrete NLoS} represents the Fourier series expansion of $h^{\nlos}(r,s)$ under the assumption that the normalized lengths at the source and receiver are large, i.e., $\frac{L_{\textrm{s}}}{\lambda}\gg1$ and $\frac{L_{\textrm{r}}}{\lambda}\gg1$. In this setting, $\sigma^2(q_x, p_x)$ is obtained as the sampled value of the \ac{PSD} at point $\big(\frac{2\pi}{L_{\textrm{r}}}q_x, \frac{2\pi}{L_{\textrm{s}}}p_x\big)$ in the wavenumber domain \cite[Eq.~(177)]{van2004detection}. We recall that, for separable scattering at the source and receiver, we have $\sigma^2(q_x, p_x)=\sigma_{\textrm{s}}^2(p_x)\sigma_{\textrm{r}}^2(q_x)$. Hence, we model only the variance at the receiver in the following, as the procedure for the source involves similar~steps.

The variance at the receiver is given by
\begin{align}
    \label{eqn: sigma at receiver}
    \sigma_{\textrm{r}}^2(q_x)\simeq\frac{1}{2\pi}S_{\textrm{r}}\bigg(\frac{2\pi}{L_{\textrm{r}}}q_x\bigg),
\end{align}
where we note that the \ac{PSD} in \eqref{eqn: Fourier Transform2} exhibits singularities due to the presence of $\sqrt{k^2-k_x^2}$ in the denominator. Therefore, rather than sampling it directly, we evaluate the integral at the nearest wavenumber-domain point as
\begin{align}
    \label{eqn: sigma at receiver1}
    \sigma_{\textrm{r}}^2(q_x)= \frac{1}{2\pi}\int_{\setW_{\textrm{r}}(q_x)}S_{\textrm{r}}\left(k_x\right)\diff k_x,
\end{align}
with $\setW_{\textrm{r}}(q_x) = \big\{\frac{2\pi}{L_{\textrm{r}}}q_x, \frac{2\pi}{L_{\textrm{r}}}(q_x+1)\big\}$.
Plugging \eqref{eqn: Fourier Transform2} into \eqref{eqn: sigma at receiver1} and  performing a change of variable similar to \eqref{eqn: autocorrelation2}, we obtain
\begin{align}
    \label{eqn: sigma at receiver2} \sigma_{\textrm{r}}^2(q_x) = \frac{1}{2\pi}\int_{{\setT}_{\textrm{r}}(q_x)}A^2_{\textrm{r}}(\theta_{\textrm{r}})\diff\theta_{\textrm{r}},
\end{align}
with $\setT_{\textrm{r}}(q_x) = \big\{\arccos\big({\frac{\lambda}{L_{\textrm{r}}}(q_{x}+1)}\big), \arccos\big({\frac{\lambda}{L_{\textrm{r}}}q_{x}}\big)\big\}$. The \ac{PSF} is determined by the function $\tilde{A}_{\textrm{r}}^2(\theta_{\textrm{r}})=\frac{{A}_{\textrm{r}}^2(\theta_{\textrm{r}})}{2\pi}$, which is modeled as a mixture of \ac{2D} distributions \cite{pizzo2022spatial}, i.e.,
\begin{align}
\label{eqn:PowerSpecralFactorLine} \tilde{A}_{\textrm{r}}^2(\theta_{\textrm{r}})=\sum_{\ell=1}^{C} w_{\ell} p_{\ell} (\theta_{\textrm{r}}),
\end{align}
where $C$ denotes the number of scattering clusters, $p_{\ell} (\theta_{\textrm{r}})$ is the \ac{2D} distribution for the $\ell$-th cluster, and the positive weights are such that $\sum_{\ell} w_{\ell}=1$. We primarily focus on the \ac{vMF} distribution, although the wrapped Gaussian distribution is also considered in the numerical results.\footnote{The \ac{2D} wrapped Gaussian distribution is defined as $ p_{\ell} (\theta_{\textrm{r}})=\frac{1}{\sqrt{2 \pi \nu_{\ell}^2}}\sum_{\tau=-\infty}^ {\infty}  \exp\big( \frac{-(\theta_{\textrm{r}} - \bar{\theta}_{\textrm{r},\ell}  + 2\pi\tau)}{2\nu_{\ell}^2}\big)$, with $\theta_{\textrm{r}}\in[-\pi, \pi)$. For simulation purposes, we retain only the term corresponding to $\tau=0$ and apply appropriate normalization to ensure the probability density function condition.} Let $I_0(\cdot)$ and $I_1(\cdot)$ denote the modified Bessel functions of the first kind and order zero and one, respectively. The \ac{2D} \ac{vMF} distribution is defined as~\cite{mardia2000directional}
\begin{align}
    \label{eqn:2DdvMF}
   \hspace{-2.5mm} p_{\ell}(\theta_{\textrm{r}}) \! = \! \frac{1}{2\pi I_0(\alpha_{\ell})}\exp\big(\alpha_{\ell}\cos{(\theta_{\textrm{r}} \! - \! \bar{\theta}_{\textrm{r},\ell}})\big), \  \theta_{\textrm{r}}\in[-\pi, \pi),
\end{align}
where $\alpha_{\ell}\geq 0$ is the concentration parameter and $\bar{\theta}_{\textrm{r},\ell}\in[-\pi, \pi)$ is the mean angle of the $\ell$-th cluster. For a given normalized circular variance $\nu_{\ell}^2\in(0, 1]$,\footnote{For measurement-based channel models, a more practically significant parameter is the angular spread, which is linked to the circular variance $\nu_{\ell}$ through the relation $\sqrt{-2\ln(1-\nu_{\ell}^2)}$ \cite{mardia2000directional}. The mean angles, angular spreads, and number of scattering clusters can be estimated or mapped from channel measurements, as detailed in the standardized channel model 3GPP TR~38.901 \cite{3gpp38901v1700}.}\label{fn:channel_measurements} we compute $\alpha_{\ell}$ using the fixed-point equation $\nu_{\ell}^2=1-\big(\frac{I_1(\alpha_{\ell})}{I_{0}(\alpha_{\ell})}\big)^2$. Note that setting $C=1$ and $\alpha_{1}=0$, which implies $\nu_{1}^2=1$, yields the case of isotropic scattering.

Having characterized the variance, the \ac{WDM}-applied \ac{NLoS} channel matrix can be obtained as
\begin{align}
    \label{eqn: Complete channel matrix}
    \bar{\H}^{\nlos}=\bar{\R}_{\textrm{r}}^{\frac{1}{2}} \W \bar{\R}_{\textrm{s}}^{\frac{1}{2}},
\end{align}
where $\bar{\R}_{\textrm{s}}$ and $\bar{\R}_{\textrm{r}}$ are defined in \eqref{eqn: Autocorrelation Matrix Source Diff and Receiver Diff} and $\W\in \mathbb{C}^{n_{\textrm{r}} \times n_{\textrm{s}}}$ is a random matrix with \ac{i.i.d.} $\setN_{\mathbb{C}}(0,1)$ entries (cf. \eqref{eqn:AngularChannelMatrixStochastic}). Furthermore, under the assumption of separable scattering, \eqref{eqn:AngularChannelMatrixStochastic} reduces to \cite{pizzo2022fourier}
\begin{align}
\label{eqn:AngularChannelMatrixStochasticSeparable}
\H_{\textrm{a}}=\diag(\sigmab_{\textrm{r}})\W\diag(\sigmab_{\textrm{s}}).
\end{align}
Finally, substituting \eqref{eqn:AngularChannelMatrixStochasticSeparable} into \eqref{eqn:ChannelMatrixStochastic}, the resulting \ac{NLoS} channel matrix is expressed as
\begin{align}
        \label{eqn:ChannelMatrixStochasticSeparable}
        \H^{\nlos}=\,& \big(\A_{\textrm{r}} \diag(\sigmab_{\textrm{r}})e^{j\diag(\gammab_{\textrm{r}})r_z}\big)\W\nonumber\\&\times\big(e^{-j\diag(\gammab_{\textrm{s}})s_z}\diag(\sigmab_{\textrm{s}})\A_{\textrm{s}}^\herm\big).
\end{align}

\subsection{Closed-Form Expression for the ACF and PSD}
\label{sec: Numerical Analysis}

We now derive closed-form expressions for the \ac{ACF} and \ac{PSD}. For simplicity, we consider a single scattering cluster; for multiple clusters, the overall \ac{ACF} and \ac{PSD} are obtained as the weighted sums of the individual contributions. Using \eqref{eqn:PowerSpecralFactorLine}--\eqref{eqn:2DdvMF} with $C=1$ and omitting the cluster index, we have
\begin{align}
    \label{eqn: one cluster psf}
   \hspace{-2mm}\tilde{A}_{\textrm{r}}^2(\theta_{\textrm{r}})=\frac{1}{2\pi I_0(\alpha)} \exp\big(\alpha\cos{(\theta_{\textrm{r}}-\bar{\theta}_{\textrm{r}})}\big), \  \theta_{\textrm{r}}\in[-\pi, \pi).
\end{align}
We only consider the forward-traveling wave, with $\theta_{\textrm{r}}\in[0, \pi)$ and $\bar{\theta}_{\textrm{r}}\in[0, \pi)$. Since \eqref{eqn: one cluster psf} is highly concentrated around the mean, extending the integration domain in \eqref{eqn: autocorrelation2} from $[0,\pi]$ to $[-\pi,\pi)$ allows to obtain an approximate yet accurate closed-form expression for the \ac{ACF}, with negligible impact on its numerical value. Therefore, plugging \eqref{eqn: one cluster psf} into \eqref{eqn: autocorrelation2} and following the approach of \cite{abdi2002parametric} yields
\begin{align}
    \label{eqn: pho one cluster}
   \Gamma_{\textrm{r}}(r_x)\simeq\frac{ I_0\big(\sqrt{\alpha^2-k^2 r_x^2+2j\alpha r_x\cos\bar{\theta}_{\textrm{r}}} \big)}{I_0(\alpha)} .
\end{align}
Taking the Fourier transform of \eqref{eqn: pho one cluster}, and utilizing \eqref{eqn: Fourier Transform2}, \eqref{eqn: one cluster psf}, and the relation $\cos\theta_{\textrm{r}} = \frac{k_x}{k}$, we have
\begin{align}
    \label{eqn: psd one cluster}
   S_{\textrm{r}}(k_x)\simeq \,& \frac{2}{\sqrt{k^2-k_x^2}}\frac{\exp\left(\alpha \cos{\bar{\theta}_{\textrm{r}}}\frac{k_x}{k}\right)}{I_0(\alpha)}\nonumber \\
   &\times\exp\bigg(\alpha\sin\bar{\theta}_{\textrm{r}}\sqrt{1-\left(\frac{k_x}{k}\right)^2}\bigg), \  |k_x|\leq k.
\end{align}

For isotropic scattering, \eqref{eqn: one cluster psf} simplifies to
\begin{align}
    \label{eqn: one isotropic}
\tilde{A}_{\textrm{r}}^2(\theta_{\textrm{r}})=\frac{1}{\pi}, \  \theta_{\textrm{r}}\in[0, \pi).
\end{align}
In this case, since \eqref{eqn: one isotropic} no longer depends on $\bar{\theta}_{\textrm{r}}$, \eqref{eqn: autocorrelation2} can be expressed as $\Gamma_{\textrm{r}}(r_x)={J_0(k r_x)}$ \cite[Eq.~(9.1.21)]{abramowitz1972handbook}, where $J_0(\cdot)$ is the Bessel function of the first kind and order zero, and the corresponding \ac{PSD} is given by $S_{\textrm{r}}(k_x)=\frac{2}{\sqrt{k^2-k_x^2}}$, $|k_x|\leq k$. Hence, under isotropic scattering, the resulting \ac{ACF} and \ac{PSD} recover the classical Jakes’ isotropic model \cite{goldsmith2005wireless}.

\section{Numerical Results}
\label{sec: Numerical Results}

In this section, we first outline the considered communication metrics and then present the performance evaluation of the \ac{EM}-based and \ac{WDM}-applied \ac{LoS}+\ac{NLoS} channels analyzed in Sections~\ref{sec:holographicMIMO}--\ref{sec: Calculation of Channel Coefficients for WDM and holographic channel}. We note that the \ac{EM}-based and \ac{WDM}-applied models refer to the spatial-sampling-based and expansion-based formulations, respectively.

\subsection{Performance Metrics}
\label{sec: Performance Metrics}

\textbf{\textit{DoF.}} In a purely \ac{LoS} scenario, the \ac{DoF} are defined as
\cite{ruiz2023degrees} 
\begin{align}
\label{eqn:DoF}
\textrm{DoF}_{\los}=\left\lfloor\frac{L_{\textrm{s}}L_{\textrm{r}}}{\lambda d}\right\rfloor.
\end{align}
In a purely \ac{NLoS} scenario with isotropic scattering, the \ac{DoF} are defined as the minimum number of non-zero coupling coefficients in $\setE_{\textrm{s}}$ and $\setE_{\textrm{r}}$ required to represent $h^{\nlos}(\r,\s)$ over the linear regions $\setL_{\textrm{s}}$ and $\setL_{\textrm{r}}$ \cite{pizzo2020degrees}, i.e.,
\begin{align}
    \label{eqn: DoF isotropic}
\textrm{DoF}_\textrm{iso}=\mathrm{min}(n_{\textrm{s}}, n_{\textrm{r}}).
\end{align}
In a purely \ac{NLoS} scenario with non-isotropic scattering, the \ac{DoF} coincide with the \ac{DoF} of the underlying random process \cite{franceschetti2017wave, piestun2000electromagnetic}, i.e.,
\begin{align}
    \label{eqn: DoF Non isotropic def}
    \textrm{DoF}_\textrm{non-iso} \! = \! \min \biggl\{ n_{\textrm{s}}' : \sum_{i=1}^{n_{\textrm{s}}'} \sigma_{\textrm{s},i}^{2} \! \geq \! 1 \! - \! \epsilon, \, n_{\textrm{r}}' : \sum_{i=1}^{n_{\textrm{r}}'} \sigma_{\textrm{r},i}^{2} \! \geq \! 1 \! - \! \epsilon \biggr\},
\end{align}
where $\epsilon$ specifies the desired level of accuracy and $\{\sigma_{\textrm{s},i}^{2}\}_{i=1}^{n_{\textrm{s}}}$ and $\{\sigma_{\textrm{r},i}^{2}\}_{i=1}^{n_{\textrm{r}}}$ represent $\big\{\sigma_{\textrm{s}}^{2}(p_x): p_x \in \setE_{\textrm{s}}\big\}$ and $\big\{\sigma_{\textrm{r}}^{2}(q_x): q_x \in \setE_{\textrm{r}}\big\}$, respectively, obtained as in \eqref{eqn: sigma at receiver1} and sorted in decreasing order.\footnote{Due to the assumption of separable scattering, $\textrm{DoF}_\textrm{non-iso}$ in \eqref{eqn: DoF Non isotropic def} depends on separate variance profiles at the source and receiver, and thus on $n_{\textrm{s}}+n_{\textrm{r}}$ independent variance elements. Under non-separable scattering, the definition would change to $\textrm{DoF}_\textrm{non-iso} = \min \big\{ n': \sum_{i=1}^{n'} \sigma_{i}^{2} \geq 1-\epsilon \big\}$, where $\{\sigma_{i}^{2}\}_{i=1}^{n_{\textrm{s}}n_{\textrm{r}}}$ represent $\big\{\sigma^{2}(q_x, p_x): (q_x, p_x) \in \setE_{\textrm{r}}\Cross\setE_{\textrm{s}}\big\}$. This definition depends on the joint variance profile and thus on $n_{\textrm{s}}n_{\textrm{r}}$ independent variance elements. Consequently, non-separable scattering is expected to result in a larger $\textrm{DoF}_\textrm{non-iso}$.}\label{fn:DoF} Lastly, extending the rationale of \eqref{eqn: DoF Non isotropic def} to \eqref{eqn: General Channel Matrix}, the \ac{DoF} in a \ac{LoS}+\ac{NLoS} scenario are obtained from the eigenvalues of $\Exp[\H^{\herm}\H] = (\H^{\los})^{\herm}\H^{\los} + \tr(\R_{\textrm{r}})\R_{\textrm{s}}$ and $\Exp[\H\H^{\herm}] = \H^{\los}(\H^{\los})^{\herm} + \tr(\R_{\textrm{s}})\R_{\textrm{r}}$ as
\begin{align}
\label{eqn: DoF def}
    \textrm{DoF}_\textrm{LoS+NLoS} =\,& \min \biggl\{ n_{\textrm{s}}' : \sum_{i=1}^{n_{\textrm{s}}'} \varrho_{i} \bigg( \frac{\Exp [\H^{\herm} \H]}{\tr \big( \Exp [\H^{\herm} \H] \big)} \bigg) \geq 1 - \epsilon, \nonumber\\&  n_{\textrm{r}}' :\sum_{i=1}^{n_{\textrm{r}}'} \varrho_{i} \bigg( \frac{\Exp [\H \H^{\herm}]}{\tr \big( \Exp [\H \H^{\herm}] \big)} \bigg) \geq 1 - \epsilon \biggr\},
\end{align}
where $\varrho_{i}(\cdot)$ denotes the $i$-th eigenvalue of the matrix argument sorted in decreasing order. Note that the \ac{DoF} definition in \eqref{eqn: DoF Non isotropic def} can be recovered as a special case of \eqref{eqn: DoF def}. In Section~\ref{sec: Numerical Results1}, as done in \cite{pizzo2022fourier}, we set $\epsilon = 0.3\%$ according to the three-sigma rule of the Gaussian distribution, stating that about $99.7\%$ of the values lie within three standard deviations.

\smallskip

\textbf{\textit{Ergodic capacity.}} Considering \eqref{eqn: received signal} and assuming perfect channel state information at both the source and receiver, the ergodic capacity (measured in bits/s/Hz) is given by \cite{tse2004fundamentals}
\begin{align}
    \label{eqn: Spectral Efficiency CSIT}  
    C_{\H}
    = \Exp \bigg[ \sum_{i=1}^{\mathrm{rank}(\H)}
      \log_{2} \left(1+\frac{P_{i}}{\chi^{2}}
      \varrho_{i}(\H\H^{\herm})\right) \bigg],
\end{align}
where $P_i$ is the transmit power allocated to the $i$-th eigenmode of $\H\H^{\herm}$, computed via water-filling for each channel realization, and $P = \sum_{i=1}^{\mathrm{rank}(\H)} P_i$ represents the total transmit power. Since deriving a closed-form characterization of the ergodic capacity remains an open problem, even for the standard \ac{MIMO} case, we evaluate \eqref{eqn: Spectral Efficiency CSIT} numerically in Section~\ref{sec: Numerical Results1} by averaging over $500$ independent channel realizations. For simplicity, we set the \ac{AWGN} variance to $\chi^2=0$~dBW and only vary $P$.

\subsection{Performance Evaluation}
\label{sec: Numerical Results1}

We consider holographic lines separated by a distance $d=10$~m, with lengths $L_{\textrm{s}} = L_{\textrm{r}}=128\lambda$ and wavelength $\lambda=0.01$~m (corresponding to a carrier frequency of $30$~GHz). Unless otherwise stated, the spatial sampling spacings at the source and receiver are set to $\Delta_\textrm{s}=\Delta_\textrm{r}=\frac{\lambda}{2}$, resulting in $N_{\textrm{r}}=N_{\textrm{s}}=256$.

\begin{figure}[t]
\centering
    \pgfdeclarelayer{background}
\pgfdeclarelayer{foreground}
\pgfsetlayers{background,main,foreground}
\begin{tikzpicture}
\begin{pgfonlayer}{background}
   \begin{axis}[
    clip=false,
	width=8.5cm,
	height=5.8cm,
	xmin=1, xmax=256,
	ymin=-200, ymax=0,
	xlabel={Eigenvalue index},
	ylabel={Normalized eigenvalue [dB]},
        xtick={1, 32, 64, ..., 256},
	xlabel near ticks,
	ylabel near ticks,
	x label style={font=\footnotesize},
	y label style={font=\footnotesize},
	ticklabel style={font=\footnotesize},
	legend style={at={(0.99,0.99)}, anchor=north east},
	legend style={font=\scriptsize, inner sep=1pt, fill opacity=0.75, draw opacity=1, text opacity=1},
	legend cell align=left,
	grid=both,
	title style={font=\scriptsize},
]

\draw[very thick, black, densely dotted] (axis cs:16,0) -- (axis cs:16,-200);
\node[black, font=\scriptsize, anchor=west] at (axis cs:16,-185) {$\textrm{DoF}_{\los} = 16$};

\addplot[thick, black] 
table [x=Var1, y=Eigen_Values_dB_rayt, col sep=comma] {Results/files_txt/LoS_SingularValues_normalized_with_trace.txt};
\addlegendentry{Ray-tracing-based}

\addplot[thick, red, only marks, mark=x, mark repeat=3]
table [x=Var1, y=Eigen_Values_dB_angular2, col sep=comma] {Results/files_txt/LoS_SingularValues_normalized_with_trace.txt};
\addlegendentry{\acs{EM}-based}

\addplot[thick, blue, only marks, mark=triangle, mark repeat=3, restrict x to domain=1:25]
table [x=Var1, y=Eigen_Values_dB_WDM_Math, col sep=comma] {Results/files_txt/LoS_SingularValues_normalized_with_trace.txt};
\addlegendentry{\acs{WDM}-applied}

\addplot[thick, green, dashed, restrict x to domain=1:25]
table [x=Var1, y=Eigen_Values_dB_WDM_Math_Luca, col sep=comma] {Results/files_txt/LoS_SingularValues_normalized_with_trace.txt};
\addlegendentry{\hspace{-0.8mm} \cite[Eq.~(62)]{sanguinetti2022wavenumber}}

\coordinate (zoom_coord) at (70,-80);
\draw[black] (1,-10) rectangle (20,-32) {};
\begin{scope}[>=latex]
\draw[-] (axis cs:20,-32) -- (axis cs:72.5,-80.5) {};
\end{scope}
    \end{axis}
\end{pgfonlayer}

\begin{pgfonlayer}{foreground}
\begin{axis}[
	axis background/.style={fill=white},
	at={(zoom_coord)},
	anchor={outer north west},
	width=0.25\textwidth,
	height=0.18\textwidth,
	xmin=1, xmax=20,
    ymin=-32, ymax=-10,
	ticks=none,
]

\draw[very thick, black, densely dotted] (axis cs:16,0) -- (axis cs:16,-200);

\addplot[thick, black] 
table [x=Var1, y=Eigen_Values_dB_rayt, col sep=comma] {Results/files_txt/LoS_SingularValues_normalized_with_trace.txt};
   
\addplot[thick, red, only marks, mark=x, mark repeat=3]
table [x=Var1, y=Eigen_Values_dB_angular2, col sep=comma] {Results/files_txt/LoS_SingularValues_normalized_with_trace.txt};
    
\addplot[thick, blue, only marks, mark=triangle, mark repeat=3, restrict x to domain=1:25]
table [x=Var1, y=Eigen_Values_dB_WDM_Math, col sep=comma] {Results/files_txt/LoS_SingularValues_normalized_with_trace.txt};
    
\addplot[thick, green, dashed, restrict x to domain=1:25]
table [x=Var1, y=Eigen_Values_dB_WDM_Math_Luca, col sep=comma] {Results/files_txt/LoS_SingularValues_normalized_with_trace.txt};

\end{axis}
\end{pgfonlayer}

\end{tikzpicture}
    \caption{Purely \ac{LoS} scenario: normalized eigenvalues of $\H^\los (\H^\los)^\herm$.}
    \label{fig:Fig2}
\end{figure}

Starting with the purely \ac{LoS} scenario, Fig.~\ref{fig:Fig2} compares the eigenvalues of $\H^\los (\H^\los)^\herm$ (normalized by its trace) for three models: the ray-tracing-based and \ac{EM}-based models described in Section~\ref{sec: Line of Sight Channel Model} and given in \eqref{eqn:Ray Tracing Channel Gain} and \eqref{eqn:Closed LoS Matrix}, respectively, and the \ac{WDM}-applied \ac{LoS} model described in Section~\ref{sec: WDM for LoS} and given in \eqref{eqn: WDM LoS H_nm}. For comparison, we additionally include the expression in \cite[Eq.~(62)]{sanguinetti2022wavenumber}, which is based on the vector Green’s function. We observe that the normalized eigenvalues corresponding to all four models closely match and consistently lead to $\textrm{DoF}_{\los}=16$. The practically significant eigenvalues are highlighted in the zoomed-in window. We note that the very small eigenvalues are inherently neglected in both the capacity evaluation, due to the water-filling power allocation, and the \ac{DoF} characterization, by definition (see Section~\ref{sec: Performance Metrics}).

Moving to the purely \ac{NLoS} scenario, we assume separable and symmetric\footnote{The assumption of symmetric scattering is adopted only to simplify the performance evaluation and does not affect the generality of the definitions and analysis of this work.}\label{fn:symmetric_scattering} scattering between the source and receiver, i.e., $\tilde{A}_\textrm{s}^2(\theta_\textrm{s})=\tilde{A}_\textrm{r}^2(\theta_\textrm{r})$, which yields $\sigma_\textrm{r}^2(q_x)=\sigma_\textrm{s}^2(p_x)$. For non-isotropic scattering, we consider three settings, which are summarized in Table~\ref{tab:parameters}:
\begin{itemize}
    \item \textbf{Non-isotropic~1} (depicted in blue) considers $C=2$ scattering clusters modeled using the \ac{2D} \ac{vMF} distribution, with mean angles $\bar{\theta}_{\textrm{r},1}=30^\circ$ and $\bar{\theta}_{\textrm{r},2}=60^\circ$ and normalized circular variances $\nu_{1}^2=0.01$ and $\nu_{2}^2=0.005$, respectively. Each cluster is assigned an equal weight in \eqref{eqn:PowerSpecralFactorLine}, i.e., $w_{1} = w_{2} = \frac{1}{2}$.
    
    \item \textbf{Non-isotropic~2} (depicted in green) considers a single scattering cluster (i.e., $C=1$) modeled using the \ac{2D} \ac{vMF} distribution, with mean angle $\bar{\theta}_{\textrm{r},1}=120^\circ$ and normalized circular variance $\nu_{1}^2=0.025$.
    
    \item \textbf{Non-isotropic~3} (depicted in magenta) considers a single scattering cluster (i.e., $C=1$) modeled using the \ac{2D} wrapped Gaussian distribution, with mean angle $\bar{\theta}_{\textrm{r},1}=90^\circ$ and normalized circular variance $\nu_{1}^2=0.03$.
\end{itemize}

\begin{table}[t]
\centering
\resizebox{\columnwidth}{!}{%
\begin{tabular}{|c|c|c|c|}
\hline
\backslashbox{\textbf{Parameters}$\downarrow$}{\textbf{Setting}$\rightarrow$}
& \textbf{\textcolor{blue}{Non-isotropic~1}}
& \textbf{\textcolor{green}{Non-isotropic~2}} 
& \textbf{\textcolor{magenta}{Non-isotropic~3}} \\ \hline \hline
\textbf{Distribution of $\tilde{A}_{\textrm{r}}^2(\theta_{\textrm{r}})$} & \ac{vMF} & \ac{vMF} & Wrapped Gaussian\\ \hline
\textbf{Number of clusters} $C$  & 2 & 1 & 1 \\ \hline
\textbf{Mean angles} $\bar{\theta}_{\textrm{r},\ell}$ & $30^\circ$, $60^\circ$ & $120^\circ$ & $90^\circ$ \\ \hline
\textbf{Circular variances} $\nu_{\ell}$ &$0.01$, $0.005$  & $0.025$ & $0.03$ \\ \hline
\end{tabular}%
}
\caption{Purely \ac{NLoS} scenario: summary of the considered non-isotropic scattering settings.}
\label{tab:parameters}
\end{table}

\begin{figure}[t]
 \centering
    \begin{subfigure}{0.48\textwidth}
    \centering
    \begin{tikzpicture}
\begin{axis}[
	width=8.5cm, height=6.5cm,
	xmin=-1, xmax=1,
	ymin=0, ymax=1,
	zmin=0, zmax=3,
	view={-30}{30},
	xlabel={$x$},
	ylabel={$z$},
	zlabel={$\tilde{A}_{\textrm{r}}^2(\theta_{\textrm{r}})$},
	x label style={at={(axis description cs:0.74,0.02)}, anchor=north},
	y label style={at={(axis description cs:0.1,0.07)}, anchor=north},
	label style={font=\footnotesize},
	xtick={-1,-0.5,...,1},
	ytick={0,0.25,...,1},
	ztick={0,1,...,3},
	ticklabel style={font=\footnotesize},
	legend style={at={(0.01,0.8)}, anchor=south west, font=\scriptsize, inner sep=1pt, fill opacity=0.75, draw opacity=1, text opacity=1},
	legend cell align=left,
	title style={font=\scriptsize, yshift=-2mm},
	grid=major
]

\addplot3[thick, red] 
table [x=Var1, y=Var2, z=Var4, col sep=comma]{Results/files_txt/PAS_Iso.txt};
\addlegendentry{Isotropic}

\addplot3[thick, blue] 
table[x=Var1, y=Var2, z=Var4, col sep=comma]{Results/files_txt/PAS_NoNIso.txt};
\addlegendentry{Non-isotropic 1}

\addplot3[thick, green] 
table[x=Var1, y=Var2, z=Var4, col sep=comma]{Results/files_txt/PAS_NoNIso1.txt};
\addlegendentry{Non-isotropic 2}

\addplot3[thick, magenta] 
table[x=Var1, y=Var2, z=Var4, col sep=comma]{Results/files_txt/PAS_NoNIso2.txt};
\addlegendentry{Non-isotropic 3}

\end{axis}

\end{tikzpicture}\vspace{-1mm}
    \caption{Illustration of $\tilde{A}_\textrm{r}^2(\theta_\textrm{r})$}
    \label{fig:Fig3a}
   \end{subfigure}\\ \vspace{3mm}
   \begin{subfigure}{0.48\textwidth}
    \centering
     \begin{tikzpicture}
   \begin{axis}[
	width=8.5cm,
	height=5.8cm,
	xmin=-128, xmax=128,
	ymin=0, ymax=1,
	xlabel={$q_x$},
	ylabel={Normalized $\sigma_{\textrm{r}}^2(q_x)$ },
        xtick={-128,-96,...,128},
	xlabel near ticks,
	ylabel near ticks,
	x label style={font=\footnotesize},
	y label style={font=\footnotesize},
	ticklabel style={font=\footnotesize},
	legend style={at={(0.99,1.01)}, anchor=south east},
	legend style={font=\scriptsize, inner sep=1pt, fill opacity=0.75, draw opacity=1, text opacity=1},
	legend cell align=left,
	grid=both,
	title style={font=\scriptsize},
]  
\addplot[thick, red]
    table [x=Var1, y=proposed, col sep=comma] {Results/files_txt/Normalized_Coupling_Coefficient_Iso_without_dB.txt};
    \addlegendentry{Isotropic \eqref{eqn: sigma at receiver2}}
\addplot[thick, red, only marks, mark=o, mark repeat=3] 
    table [x=Var1, y=pizzo, col sep=comma] {Results/files_txt/Normalized_Coupling_Coefficient_Iso_without_dB.txt};
    \addlegendentry{Isotropic (\hspace{-0.8mm} \cite[Eq.~(69)]{pizzo2020spatially})}

    \addplot[thick, blue]
    table [x=Var1, y=Var2, col sep=comma] {Results/files_txt/Normalized_Coupling_Coefficient_NoNIso_without_dB.txt};
    \addlegendentry{Non-isotropic 1  \eqref{eqn: sigma at receiver2}}

    \addplot[thick, green]
    table [x=Var1, y=Var2, col sep=comma] {Results/files_txt/Normalized_Coupling_Coefficient_NoNIso_without_dB1.txt};
    \addlegendentry{Non-isotropic 2 \eqref{eqn: sigma at receiver2}}

     \addplot[thick, magenta]
    table [x=Var1, y=Var2, col sep=comma] {Results/files_txt/Normalized_Coupling_Coefficient_NoNIso_without_dB2.txt};
    \addlegendentry{Non-isotropic 3 \eqref{eqn: sigma at receiver2}}
    
    \end{axis}
\end{tikzpicture}\vspace{-1mm}
     \caption{Normalized $\sigma_{\textrm{r}}^2(q_x)$}\label{fig:Fig3b} \vspace{2mm}
\end{subfigure}
\caption{Purely \ac{NLoS} scenario: \acp{PSF} and the corresponding normalized variances for the considered isotropic and non-isotropic scattering settings, where the latter are summarized in Table~\ref{tab:parameters}.}\label{fig:Fig3}
\end{figure}
\begin{figure}[t]
\centering    
\begin{tikzpicture}[>=latex]
   \begin{axis}[
	width=8.5cm,
	height=5.8cm,
	xmin=1, xmax=256,
	ymin=-60, ymax=-10,
	xlabel={Eigenvalue index},
	ylabel={Normalized eigenvalue [dB]},
       xtick={1,32,64,...,256},
      ytick={-60,-50,...,-10},
	xlabel near ticks,
	ylabel near ticks,
	x label style={font=\footnotesize},
	y label style={font=\footnotesize},
	ticklabel style={font=\footnotesize},
	legend style={at={(0.99,0.01)}, anchor=south east},
	legend style={font=\scriptsize, inner sep=1pt, fill opacity=0.75, draw opacity=1, text opacity=1},
	legend cell align=left,
	grid=both,
	title style={font=\scriptsize},
]  
    \addplot[thick, black] 
    table [x=Var1, y=iid, col sep=comma] {Results/files_txt/Rr_Eigen_0.5lambda_normalized.txt};
    \addlegendentry{i.i.d. Rayleigh fading}
    \addplot[thick, red, only marks, mark=star, mark options=solid, mark repeat=6]
    table [x=Var1, y=Eigen_values_ClarkedB, col sep=comma] {Results/files_txt/Rr_Eigen_0.5lambda_normalized.txt};
    \addlegendentry{Jakes' isotropic}
    \addplot[thick, red]
     table [x=Var1, y=Iso, col sep=comma] {Results/files_txt/Rr_Eigen_0.5lambda_normalized.txt};
    \addlegendentry{Isotropic}   
    \addplot[thick, blue]
    table [x=Var1, y=NonIso, col sep=comma] {Results/files_txt/Rr_Eigen_0.5lambda_normalized.txt};
    \addlegendentry{Non-isotropic 1}
    \addplot[thick, green]
    table [x=Var1, y=NonIso, col sep=comma] {Results/files_txt/Rr_Eigen_0.5lambda_normalized_noniso1.txt};
    \addlegendentry{Non-isotropic 2}

    \addplot[thick, magenta]
    table [x=Var1, y=NonIso, col sep=comma] {Results/files_txt/Rr_Eigen_0.5lambda_normalized_noniso2.txt};
    \addlegendentry{Non-isotropic 3}

\draw[very thick, blue, densely dotted] (axis cs:82,-10) -- (axis cs:82,-60);
\node[black, font=\scriptsize, anchor=east] at (axis cs:82,-45) {$\textrm{DoF}_{\textrm{non-iso}} = 82$};
\draw[very thick, green, densely dotted] (axis cs:101,-10) -- (axis cs:101,-60);
\node[black, font=\scriptsize, anchor=west] at (axis cs:125,-15) {$\textrm{DoF}_{\textrm{non-iso}} = 101$};
\draw[<-, thin] (axis cs:101,-15) -- (axis cs:128,-15);

\draw[very thick, magenta, densely dotted] (axis cs:125,-10) -- (axis cs:125,-60);
\node[black, font=\scriptsize, anchor=west] at (axis cs:125,-20) {$\textrm{DoF}_{\textrm{non-iso}} = 125$};

\end{axis}
\end{tikzpicture}
 \caption{Purely \ac{NLoS} scenario: normalized eigenvalues of $\R_{\textrm{r}}$. }\label{fig:Fig6}
\end{figure}

Fig.~\ref{fig:Fig3a} illustrates $\tilde{A}_\textrm{r}^2(\theta_\textrm{r})$ for isotropic and non-isotropic scattering, obtained using \eqref{eqn: one isotropic} and \eqref{eqn:PowerSpecralFactorLine}, respectively. In all cases, only forward-traveling waves are considered by considering $\theta_{\textrm{r}} \in [0,\pi)$. The corresponding variance $\sigma_{\textrm{r}}^2(q_x)$ (normalized by the maximum value) is plotted in Fig.~\ref{fig:Fig3b}, showing the average channel power of each wavenumber component. To validate our modeling, the variance in \eqref{eqn: sigma at receiver2} is compared with \cite[Eq.~(69)]{pizzo2020spatially} in the isotropic case.

\begin{figure}[t]
\centering
    \begin{tikzpicture}
   \begin{axis}[
	width=8.5cm,
	height=5.8cm,
	xmin=2, xmax=8,
	ymin=0, ymax=6,
	xlabel={$\Delta_\textrm{s}=\Delta_\textrm{r}$},
	ylabel={$C_{\H}$ [kbit/s/Hz]},
       xtick={2,4,6,8},
      ytick={0,1,2,3,4,5,6},
      xticklabels={$\frac{\lambda}{2}$,$\frac{\lambda}{4}$,$\frac{\lambda}{6}$,$\frac{\lambda}{8}$},
	xlabel near ticks,
	ylabel near ticks,
	x label style={font=\footnotesize},
	y label style={font=\footnotesize},
	ticklabel style={font=\footnotesize},
    x dir=reverse,
	legend style={at={(0.99,0.99)}, anchor=north east},
	legend style={font=\scriptsize, inner sep=1pt, fill opacity=0.75, draw opacity=1, text opacity=1},
	legend cell align=left,
	grid=both,
	title style={font=\scriptsize},
]  
    \addplot[thick, black] 
    table [x=Spacing, y=iid, col sep=comma] {Results/files_txt/Spacing_SE_NLoS_CSIT.txt};
    \addlegendentry{i.i.d. Rayleigh fading}
    \addplot[thick, red, only marks,  mark=star, mark options=solid]
    table [x=Spacing, y=Jakes, col sep=comma] {Results/files_txt/Spacing_SE_NLoS_CSIT.txt};
    \addlegendentry{Jakes' isotropic}
    \addplot[thick, red]
     table [x=Spacing, y=Iso, col sep=comma] {Results/files_txt/Spacing_SE_NLoS_CSIT.txt};
    \addlegendentry{Isotropic}   
    \addplot[thick, blue]
    table [x=Spacing, y=NonIso1, col sep=comma] {Results/files_txt/Spacing_SE_NLoS_CSIT.txt};
    \addlegendentry{Non-isotropic 1}
    \addplot[thick, green]
    table [x=Spacing, y=NonIso2, col sep=comma] {Results/files_txt/Spacing_SE_NLoS_CSIT.txt};
    \addlegendentry{Non-isotropic 2}
    \addplot[thick, magenta]
    table [x=Spacing, y=NonIso3, col sep=comma] {Results/files_txt/Spacing_SE_NLoS_CSIT.txt};
    \addlegendentry{Non-isotropic 3}
    \end{axis}
\end{tikzpicture}
     \caption{Purely \ac{NLoS} scenario: ergodic capacity versus spatial sampling spacing, with $P = 20$~dBW.}\label{fig:Fig8}
\end{figure} 
\begin{figure}[t]
\centering
     \begin{tikzpicture}

   \begin{axis}[
	width=8.5cm,
	height=5.8cm,
	xmin=0, xmax=30,
	ymin=0, ymax=4,
    xlabel={$P$ [dBW]},
	ylabel={$C_{\H}$ [kbit/s/Hz]},
      ytick={0,1,2,3,4},
	xlabel near ticks,
	ylabel near ticks,
	x label style={font=\footnotesize},
	y label style={font=\footnotesize},
	ticklabel style={font=\footnotesize},
	legend style={at={(0.01,0.99)}, anchor=north west},
	legend style={font=\scriptsize, inner sep=1pt, fill opacity=0.75, draw opacity=1, text opacity=1},
	legend cell align=left,
	grid=both,
	title style={font=\scriptsize},
]  
\addplot[thick, solid, black]
    table[x=SNR,y=iid,col sep=comma]{Results/files_txt/SNR_SE_NLoS_CSIT_lambda_0.5.txt};
\addlegendentry{i.i.d.\ Rayleigh}

\addplot[thick, dashed, black, forget plot]
    table[x=SNR,y=iid,col sep=comma]{Results/files_txt/SNR_SE_NLoS_CSIT_lambda_0.25.txt};

\addplot[thick, red, only marks, mark=star, mark options=solid]
    table[x=SNR,y=Jakes,col sep=comma]{Results/files_txt/SNR_SE_NLoS_CSIT_lambda_0.5.txt};
\addlegendentry{Jakes' isotropic}

\addplot[thick, red, only marks, mark=star, mark options=solid, forget plot]
    table[x=SNR,y=Jakes,col sep=comma]{Results/files_txt/SNR_SE_NLoS_CSIT_lambda_0.25.txt};

\addplot[thick, solid, red]
    table[x=SNR,y=Iso,col sep=comma]{Results/files_txt/SNR_SE_NLoS_CSIT_lambda_0.5.txt};
\addlegendentry{Isotropic}

\addplot[thick, dashed, red, forget plot]
    table[x=SNR,y=Iso,col sep=comma]{Results/files_txt/SNR_SE_NLoS_CSIT_lambda_0.25.txt};

\addplot[thick, solid, blue]
    table[x=SNR,y=NonIso1,col sep=comma]{Results/files_txt/SNR_SE_NLoS_CSIT_lambda_0.5.txt};
\addlegendentry{Non-isotropic 1}

\addplot[thick, dashed, blue, forget plot]
    table[x=SNR,y=NonIso1,col sep=comma]{Results/files_txt/SNR_SE_NLoS_CSIT_lambda_0.25.txt};

\addplot[thick, solid, green]
    table[x=SNR,y=NonIso2,col sep=comma]{Results/files_txt/SNR_SE_NLoS_CSIT_lambda_0.5.txt};
\addlegendentry{Non-isotropic 2}

\addplot[thick, dashed, green, forget plot]
    table[x=SNR,y=NonIso2,col sep=comma]{Results/files_txt/SNR_SE_NLoS_CSIT_lambda_0.25.txt};

\addplot[thick, solid, magenta]
    table[x=SNR,y=NonIso3,col sep=comma]{Results/files_txt/SNR_SE_NLoS_CSIT_lambda_0.5.txt};
\addlegendentry{Non-isotropic 3}

\addplot[thick, dashed, magenta]
    table[x=SNR,y=NonIso3,col sep=comma, forget plot]{Results/files_txt/SNR_SE_NLoS_CSIT_lambda_0.25.txt};
\end{axis}
\end{tikzpicture}
     \caption{Purely \ac{NLoS} scenario: ergodic capacity versus total transmit power. Solid and dashed curves correspond to $\Delta_\textrm{s}=\Delta_\textrm{r}=\frac{\lambda}{2}$ and $\Delta_\textrm{s}=\Delta_\textrm{r}=\frac{\lambda}{4}$, respectively.}\label{fig:Fig9}
\end{figure}

As discussed in Section~\ref{sec: Autocorrelation of Channel H_{nm}}, the spectra of the spatial autocorrelation matrices in \eqref{eqn: Autocorrelation Matrix Source Diff and Receiver Diff} and \eqref{eqn: Autocorrelation Matrix Source Diff and Receiver Diff Holographic}, corresponding to the \ac{WDM}-applied and \ac{EM}-based \ac{NLoS} channels, respectively, are identical up to a constant scaling and thus have the same normalized eigenvalues. Fig.~\ref{fig:Fig6} plots the eigenvalues of the receive spatial autocorrelation matrix $\R_{\textrm{r}}$ (normalized by its trace). For \ac{i.i.d.} Rayleigh fading, the channel entries are mutually independent, leading to identical eigenvalues of $\R_{\textrm{r}}$. The closest physically meaningful counterpart is Jakes' isotropic model \cite{goldsmith2005wireless}, whose spatial autocorrelation matrix is obtained from the spatial sampling of $J_0(k r_x)$. The normalized eigenvalues under isotropic scattering coincide with those of Jakes' isotropic model, whereas the non-isotropic case exhibits a significantly steeper decay, indicating stronger spatial correlation. Comparing the different non-isotropic settings, `Non-isotropic~3' shows the slowest decay among the three due to the broadest spread of $\tilde{A}_\textrm{r}^{2}(\theta_\textrm{r})$. In terms of \ac{DoF}, we obtain $\textrm{DoF}_{\textrm{iso}}=256$ for isotropic scattering, while `Non-isotropic~1', `Non-isotropic~2', and `Non-isotropic~3' yield $\textrm{DoF}_{\textrm{non-iso}}=82$, $\textrm{DoF}_{\textrm{non-iso}}=101$, and $\textrm{DoF}_{\textrm{non-iso}}=125$, respectively. Note that reducing the spatial sampling spacing below $\frac{\lambda}{2}$ does not increase the \ac{DoF}: as implied by \eqref{eqn: DoF isotropic}--\eqref{eqn: DoF Non isotropic def} and consistent with the Nyquist sampling theorem, the \ac{DoF} depend only on the lengths of the source and receiver, not on the spatial oversampling.

Fig.~\ref{fig:Fig8} depicts the ergodic capacity against the spatial sampling spacing $\Delta_\textrm{s}=\Delta_\textrm{r}$ with $P = 20$~dBW in a purely \ac{NLoS} scenario. Under \ac{i.i.d.} Rayleigh fading, the capacity grows linearly as the spacing decreases, since more spatial samples become available: this behavior is not realistic because densely spaced samples cannot remain uncorrelated. For the physically consistent channel models, the capacity grows more gradually. In fact, while reducing the spacing does not increase the \ac{DoF} (see Fig.~\ref{fig:Fig6}), it does offer additional diversity by increasing the number of spatial samples. The capacity of Jakes' isotropic model closely matches that of isotropic scattering. The non-isotropic cases yield the lowest capacity due to the stronger spatial correlation and the correspondingly smaller number of dominant eigenmodes. Fig.~\ref{fig:Fig9} shows the ergodic capacity against the total transmit power for two different spatial sampling spacings, i.e., $\Delta_\textrm{s}=\Delta_\textrm{r}=\frac{\lambda}{4}$ and $\Delta_\textrm{s}=\Delta_\textrm{r}=\frac{\lambda}{2}$. Consistent with Fig.~\ref{fig:Fig8}, Jakes’ isotropic model again aligns with the isotropic case, whereas the non-isotropic cases perform the worst across the whole total transmit power range. For $\frac{\lambda}{2}$ spatial sampling spacing, the capacity under \ac{i.i.d.} Rayleigh fading also approaches that of the isotropic and Jakes' models.

\begin{figure}[t]
\centering
    \begin{tikzpicture}
   \begin{axis}[
	width=8.5cm,
	height=5.8cm,
	xmin=1, xmax=256,
	ymin=-200, ymax=0,
	xlabel={Eigenvalue index},
	ylabel={Normalized eigenvalue [dB]},
       xtick={1,32,64,...,256},
	xlabel near ticks,
	ylabel near ticks,
	x label style={font=\footnotesize},
	y label style={font=\footnotesize},
	ticklabel style={font=\footnotesize},
	legend style={at={(0.99,0.45)}, anchor=east},
	legend style={font=\scriptsize, inner sep=1pt, fill opacity=0.75, draw opacity=1, text opacity=1},
	legend cell align=left,
	grid=both,
	title style={font=\scriptsize},
]  

    \addplot[thick, black] 
    table [x=Var1, y=Normalized_Eigen_Values_Hankel_dB, col sep=comma] {Results/files_txt/Expectation_Eigen_HLoS_NLoS_Iso_TraceNorm.txt};
    \addlegendentry{LoS}
    
    
    \addplot[thick, red]
     table [x=Var1, y=LoSNLoSEigen_dB, col sep=comma] {Results/files_txt/Expectation_Eigen_HLoS_NLoS_Iso_TraceNorm.txt};
    \addlegendentry{LoS+NLoS (Isotropic)}
    

     \addplot[thick, blue]
     table [x=Var1, y=LoSNLoSEigen_dB1, col sep=comma] {Results/files_txt/Expectation_Eigen_HLoS_NLoS_NonIso_TraceNorm.txt};
    \addlegendentry{LoS+NLoS (Non-isotropic 1)}


    \addplot[thick, green]
     table [x=Var1, y=LoSNLoSEigen_dB2, col sep=comma] {Results/files_txt/Expectation_Eigen_HLoS_NLoS_NonIso_TraceNorm.txt};
    \addlegendentry{LoS+NLoS (Non-isotropic 2)}

    \addplot[thick, magenta]
     table [x=Var1, y=LoSNLoSEigen_dB3, col sep=comma] {Results/files_txt/Expectation_Eigen_HLoS_NLoS_NonIso_TraceNorm.txt};
    \addlegendentry{LoS+NLoS (Non-isotropic 3)}

    \draw[very thick, red, densely dotted] (axis cs:18,0) -- (axis cs: 18,-200);
    \node[black, font=\scriptsize, anchor=west] at (axis cs:18,-25) {$\textrm{DoF}_{\textrm{LoS+NLoS}} = 18$ (Isotropic)};

    \draw[very thick, blue, densely dotted] (axis cs:19,0) -- (axis cs:19,-200);
    \node[black, font=\scriptsize, anchor=west] at (axis cs:19,-185) {$\textrm{DoF}_{\textrm{LoS+NLoS}} = 19$ (Non-isotropic)};
    
    \end{axis}
\end{tikzpicture}
     \caption{\ac{LoS}+\ac{NLoS} scenario: normalized eigenvalues of $\mathbb{E}\big[\H\H^\herm\big]$.}\label{fig:Fig7}
\end{figure} 
\begin{figure}[t]
\centering
\begin{tikzpicture}
   \begin{axis}[
	width=8.5cm,
	height=5.8cm,
	xmin=0, xmax=30,
	ymin=0.5, ymax=2.5,
    xlabel={$P$ [dBW]},
	ylabel={$C_{\H}$ [kbit/s/Hz]},
      ytick={0.5,1,...,2.5},
	xlabel near ticks,
	ylabel near ticks,
	x label style={font=\footnotesize},
	y label style={font=\footnotesize},
	ticklabel style={font=\footnotesize},
	legend style={at={(0.01,0.99)}, anchor=north west},
	legend style={font=\scriptsize, inner sep=1pt, fill opacity=0.75, draw opacity=1, text opacity=1},
	legend cell align=left,
	grid=both,
	title style={font=\scriptsize},
]    
    \addplot[thick, solid, black] 
    table [x=SNR, y=Cap_LoS, col sep=comma] {Results/files_txt/SNR_SE_LoSNLoS_CSIT_lambda_0.5.txt};
    \addlegendentry{LoS}
    
    \addplot[thick, solid, red]
    table [x=SNR, y=Capacity_LoSNLoS_iso, col sep=comma] {Results/files_txt/SNR_SE_LoSNLoS_CSIT_lambda_0.5.txt};
    \addlegendentry{LoS+NLoS (Isotropic)}
    

    \addplot[thick, solid, blue]
    table [x=SNR, y=Capacity_LoSNLoS_noniso1, col sep=comma] {Results/files_txt/SNR_SE_LoSNLoS_CSIT_lambda_0.5.txt};
    \addlegendentry{LoS+NLoS (Non-isotropic 1)}

    
    \addplot[thick, solid , green]
    table [x=SNR, y=Capacity_LoSNLoS_noniso2, col sep=comma] {Results/files_txt/SNR_SE_LoSNLoS_CSIT_lambda_0.5.txt};
    \addlegendentry{LoS+NLoS (Non-isotropic 2)}

    \addplot[thick, solid , magenta]
    table [x=SNR, y=Capacity_LoSNLoS_noniso3, col sep=comma] {Results/files_txt/SNR_SE_LoSNLoS_CSIT_lambda_0.5.txt};
    \addlegendentry{LoS+NLoS (Non-isotropic 3)}
    \end{axis}
\end{tikzpicture}
\caption{\ac{LoS}+\ac{NLoS} scenario: ergodic capacity versus total transmit power.}\label{fig:Fig10}
\end{figure}

Lastly, we examine the \ac{LoS}+\ac{NLoS} scenario, where the \ac{LoS} and \ac{NLoS} components combine additively (as described in Section~\ref{sec: LoS+NLoS Channel Model}) and the latter is modeled as in Fig.~\ref{fig:Fig3}. Fig.~\ref{fig:Fig7} plots the eigenvalues of $\mathbb{E}\big[\H\H^{\herm}\big]$ (normalized by its trace) for both the purely \ac{LoS} and \ac{LoS}+\ac{NLoS} scenarios. Under isotropic scattering, adding the \ac{NLoS} component strengthens all the weaker eigenmodes of the \ac{LoS} channel. In contrast, under non-isotropic scattering, adding the \ac{NLoS} component significantly boosts only the eigenmodes associated with the directions around the scattering clusters, while the remaining eigenmodes follow the trend of the \ac{LoS} channel. In terms of \ac{DoF}, we have $\textrm{DoF}_\textrm{LoS+NLoS}=18$ for isotropic scattering and $\textrm{DoF}_\textrm{LoS+NLoS}=19$ for all the non-isotropic settings. Compared with the purely \ac{NLoS} scenarios, the reduction in the \ac{DoF} for the \ac{LoS}+\ac{NLoS} case follows from \eqref{eqn: DoF def} due to the shift in the eigenvalue distribution. The corresponding ergodic capacity against the total transmit power is illustrated in Fig.~\ref{fig:Fig10}. With water-filling power allocation, the capacity reflects the eigenvalue distribution of $\mathbb{E}\big[\H\H^{\herm}\big]$. Under isotropic scattering, the eigenvalues are more evenly balanced and the transmit power is spread across a larger set of comparable eigenmodes, yielding higher capacity at high transmit power. Under non-isotropic scattering, the more unevenly balanced eigenvalues result in lower capacity. This effect is even more pronounced in the purely \ac{LoS} scenario, where the eigenvalue distribution is most uneven.

\section{Conclusions and Future Directions}
\label{sec: Conclusions}

We analyzed the \ac{EM}-based and \ac{WDM}-applied \ac{LoS}+\ac{NLoS} channels for holographic lines. We showed that these formulations represent two sides of the same coin, as they yield equivalent eigenvalue spectra for both the \ac{LoS} and \ac{NLoS} components, differing only by a constant scaling. Furthermore, applying \ac{WDM} to the \ac{NLoS} component leads to its angular-domain representation and also enables the analytical characterization of \ac{PSD}-\ac{PSF} relationship. This demonstrates that \ac{WDM} can be seen as a powerful analytical tool for studying holographic \ac{MIMO} channels. We further derived closed-form expressions for the \ac{ACF} and \ac{PSD} under both isotropic and non-isotropic scattering, and specializing the analysis to isotropic scattering recovers Jakes’ isotropic model. Numerical results showed that incorporating the \ac{NLoS} component results in substantial performance gains relative to the purely \ac{LoS} channel.

The analysis presented in this work is derived under commonly adopted assumptions, particularly separable scattering, which make the analysis tractable while limiting the generality of the current framework. Future work will therefore focus on relaxing these assumptions and evaluating how the resulting performance differs from the findings reported in this study. In addition, future research may extend the proposed \ac{WDM} framework to multi-user scenarios and to basis functions beyond Fourier basis functions.


\appendices

\section{Comparison of Different Green's Functions}
\label{sec: Different type of Green functions and their approximation}

\begin{figure}[t]
\centering
\begin{tikzpicture}
   \begin{axis}[
	width=8.5cm,
	height=5.8cm,
	xmin=-10.24, xmax=10.24,
	ymin=-28, ymax=-20,
	xlabel={$s_x$},
	ylabel={ Amplitude [dB]},
    xtick={-10.24,-5.12,...,10.24},
	ytick={-20,-22,...,-28},
    xticklabels={$-1024\lambda$, $-512\lambda$, $0$, $512\lambda$, $1024\lambda$},
	xlabel near ticks,
	ylabel near ticks,
	x label style={font=\footnotesize},
	y label style={font=\footnotesize},
	ticklabel style={font=\footnotesize},
	legend style={at={(0.5,0.01)}, anchor=south},
	legend style={font=\scriptsize, inner sep=1pt, fill opacity=0.75, draw opacity=1, text opacity=1},
	legend cell align=left,
	grid=both,
	title style={font=\scriptsize},
]
     \addplot [
        fill=gray,
        fill opacity=0.25,
        draw=none
    ] coordinates {
        (-0.64,-28) (0.64,-28) (0.64,-20) (-0.64,-20)
    };
    \addplot[thick, black] 
    table [x=z, y=10, col sep=tab] {Results/files_txt/VectorGreen.txt};
    \addlegendentry{Vector Green's function}

    \addplot[thick, black, dashed]
    table [x=z, y=10, col sep=tab] {Results/files_txt/ScalarGreen.txt};
    \addlegendentry{Scalar Green's function}

    \addplot[thick, black, dotted] 
    table [x=z, y=10, col sep=tab] {Results/files_txt/ParaxialGreen.txt};
    \addlegendentry{Paraxial Green's function}

    \addplot[thick, black]
    table [x=z, y=25, col sep=tab] {Results/files_txt/VectorGreen.txt};
    
    \addplot[thick, black, dashed]
    table [x=z, y=25, col sep=tab] {Results/files_txt/ScalarGreen.txt};

    \addplot[thick, black, dotted]
    table [x=z, y=25, col sep=tab] {Results/files_txt/ParaxialGreen.txt};

\draw (3.20,-21.35) ellipse (0.4 and 0.6);
\draw (3.20,-25.05) ellipse (0.4 and 0.3);

\node[black, font=\scriptsize, anchor=west] at (2.65,-20.4) {$d = 10$~m};
\node[black, font=\scriptsize, anchor=west] at (2.65,-24.4) {$d = 25$~m};

\end{axis}
\end{tikzpicture}
\caption{Amplitude profiles obtained with different Green's functions, with source at $\s=[s_x, 0]^{\tran}$ and receiver at $\r=[0, d]^{\tran}$.}
\label{fig: vector green function}
\end{figure}

Here, we compare the amplitude profiles of the vector Green's function used in \cite{sanguinetti2022wavenumber}, the scalar Green's function, and the paraxial Green's function. For the system model in Fig.~\ref{fig:Fig1} with $\s=[s_x, 0]^{\tran}$ and $\r=[0, d]^{\tran}$, these are given by
\begin{subequations}
\label{eqn: Different Greens functions}
\begin{align}
\hspace{-2.5mm} G_{\text{vec}}(s_x, d) & = \frac{d^2}{4\pi}\frac{e^{jk\sqrt{s_x^2+d^2}}}{(s_x^2+d^2)^{3/2}} && \hspace{-1.5mm} \textrm{\footnotesize (vector Green's function)}, \label{eq:greens_vec} \\
\hspace{-2.5mm} G_{\text{sca}}(s_x, d) & = \frac{1}{4\pi}\frac{e^{jk\sqrt{s_x^2+d^2}}}{\sqrt{s_x^2+d^2}} && \hspace{-1.5mm} \textrm{\footnotesize (scalar Green's function)}, \label{eq:greens_sca} \\
\hspace{-2.5mm} G_{\text{par}}(s_x, d) & = \frac{1}{4\pi}\frac{e^{j\kappa(d+\frac{s_x^2}{2d})}}{d} && \hspace{-1.5mm} \textrm{\footnotesize (paraxial Green's function)}. \label{eq:greens_par}
\end{align}
\end{subequations}
Specifically, as done in \cite[Eq.~(56)]{sanguinetti2022wavenumber}, \eqref{eq:greens_vec} is obtained as the $(1,1)$-th entry of the vector Green's function, which is the component relevant to our setting. For a fair comparison, \eqref{eq:greens_sca} is derived as in Section~\ref{sec:line} by setting the $y$-coordinate to zero. Lastly, \eqref{eq:greens_par} corresponds to the paraxial approximation of \eqref{eq:greens_sca} valid for $d \gg s_x$. Fig.~\ref{fig: vector green function} shows that, even for a small distance such as $d=10$~m, the amplitude profiles of the different Green's functions remain nearly identical for $|s_x|\leq 128\lambda$ (corresponding to $L_\textrm{s}$ up to $256\lambda$). In our performance evaluation in Section~\ref{sec: Numerical Results1}, we adopt $L_\textrm{s} = 128 \lambda$, i.e., $|s_x|\leq 64\lambda$, which is highlighted by the shaded region in the figure.

\addcontentsline{toc}{chapter}{References} 
\bibliographystyle{IEEEtran}
\bibliography{refs_abbr,refs}

\begin{IEEEbiography}
[{\includegraphics[width=1in,height=1.25in,clip,keepaspectratio]{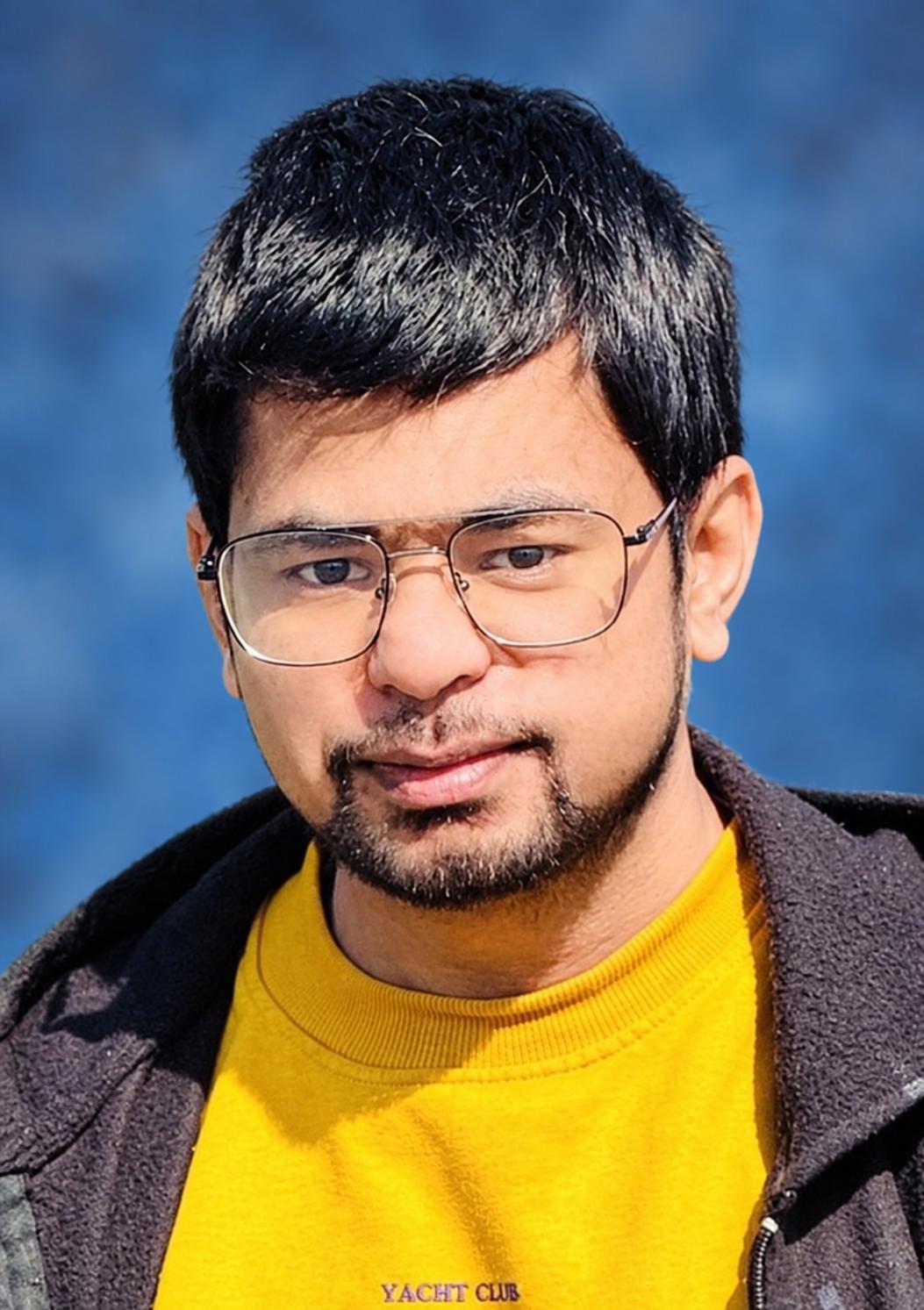}}]
{Ashutosh Prajapati} received the B.Tech. degree in electronics and communication engineering from the Indian Institute of Information Technology Sonepat (mentored by the National Institute of Technology Kurukshetra), India, in 2018, and the M.Tech. degree in communication and signal processing from the Indian Institute of Technology Ropar, India, in 2020. Since 2023, he has been a Doctoral Researcher with the Centre for Wireless Communications, University of Oulu, Oulu, Finland. From 2020 to 2023, he was a Researcher with TCS Research, India. His primary research interests include communication theory, signal processing, convex optimization, electromagnetic theory, and their applications to multi-antenna communications. He received the Institute’s Silver Medal in recognition of having the highest grade point average in the M.Tech. program. He currently serves as a reviewer for several IEEE journals.
\end{IEEEbiography}

\begin{IEEEbiographynophoto}{Prathapasinghe Dharmawansa}
received the B.Sc. and M.Sc. degrees in electronic and telecommunication
engineering from the University of Moratuwa, Sri Lanka, in 2003 and 2004,
respectively, and the D.Eng. degree in information and communication
technology from the Asian Institute of Technology, Thailand, in 2007.
He is currently a Senior Research Fellow with the Center for Wireless
Communications, University of Oulu, Finland. His research interests include
communication theory, wireless communications, signal processing, random matrix theory, and high
dimensional statistics.
\end{IEEEbiographynophoto}

\begin{IEEEbiography}[{\includegraphics[width=1in,height=1.25in,clip,keepaspectratio]{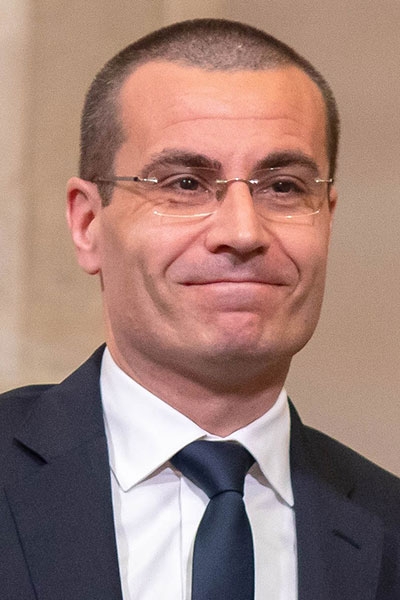}}]
{Marco Di Renzo} (Fellow, IEEE) received the Laurea (cum laude) and Ph.D. degrees in electrical engineering from the University of L’Aquila, Italy, in 2003 and 2007, respectively, and the Habilitation à Diriger des Recherches (Doctor of Science) degree from University Paris-Sud (currently Paris-Saclay University), France, in 2013. Currently, he is Chair Professor of Telecommunications Engineering, the Director of the Centre for Telecommunications Research, and the Head of the Telecommunications Group, Department of Engineering, King’s College London, London, United Kingdom. He is also a CNRS Research Director (Professor) with the Institute of Electronics and Digital Technologies (IETR) at CNRS \& CentraleSup\'elec - Univ. Paris-Saclay, Rennes, France. He was a France-Nokia Chair of Excellence in ICT at the University of Oulu (Finland), a Tan Chin Tuan Exchange Fellow in Engineering at Nanyang Technological University (Singapore), a Fulbright Fellow at The City University of New York (USA), a Nokia Foundation Visiting Professor at Aalto University (Finland), and a Royal Academy of Engineering Distinguished Visiting Fellow at Queen’s University Belfast (U.K.). He is a Fellow of the IEEE, IET, EURASIP, RSA, and AAIA; an Academician of AIIA; an Ordinary Member of the European Academy of Sciences and Arts, an Ordinary Member of the Academia Europaea, and an Ordinary Member of the Italian Academy of Technology and Engineering; an Ambassador of the European Association on Antennas and Propagation; and a Highly Cited Researcher. He has received several distinctions, including the Michel Monpetit Prize conferred by the French Academy of Sciences, the IEEE Communications Society Heinrich Hertz Award, and the IEEE Communications Society Marconi Prize Paper Award in Wireless Communications. Also, he is a principal investigator of an ERC Synergy grant on metasurface-based information processing. He served as the Editor-in-Chief of IEEE Communications Letters from 2019 to 2023, and as the Director of Journals and Chair of the Publications Misconduct Ad Hoc Committee of the IEEE Communications Society from 2024 to 2025. Currently, he sits on the IEEE-COMSOC Fellow Evaluation Standing Committee and on the Editorial Board of the Proceedings of the IEEE.
\end{IEEEbiography}

\begin{IEEEbiography}[{\includegraphics[width=1in,height=1.25in,clip,keepaspectratio]{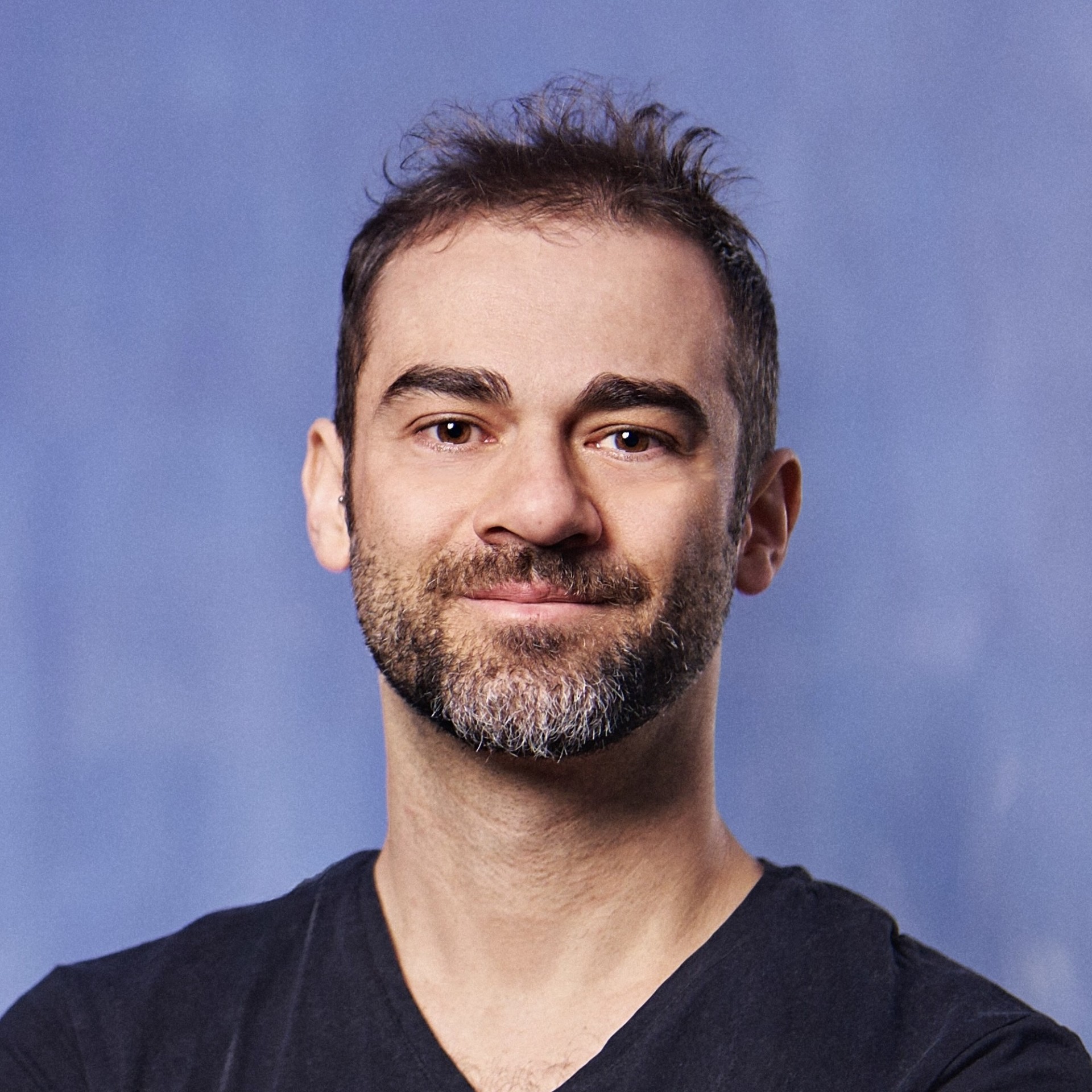}}]
{Italo Atzeni} received the MSc degree (Hons.) in telecommunications engineering from the University of Cagliari, Italy, in 2010 and the PhD degree (Hons.) in signal theory and communications from the Polytechnic University of Catalonia--BarcelonaTech, Spain, in 2014. Since 2019, he has been with the Centre for Wireless Communications, University of Oulu, Finland, where he is currently an Associate Professor and Academy Research Fellow. In 2014--2017, he was a Researcher at the Mathematical and Algorithmic Sciences Lab, Huawei Technologies, France. In 2017--2018, he was a Postdoctoral researcher at EURECOM, France. In 2013, he was a visiting doctoral research at The Hong Kong University of Science and Technology, Hong Kong. His research interests include signal processing, communication theory, convex and distributed optimization, and their applications to low-complexity, energy-efficient multi-antenna communications. He received Best Paper Awards at IEEE ICC 2019 (Wireless Communications Symposium) and IEEE SPAWC 2023. He was awarded the MSCA Individual Fellowship (2020--2022) by the European Commission and the Academy Research Fellowship (2022--2027) by the Research Council of Finland. He currently serves as an Editor for IEEE Transactions on Wireless Communications and as a Senior Editor for IEEE Communications Letters.
\end{IEEEbiography}

\end{document}